\documentclass[twocolumn,eqsecnum,showpacs,aps,prb]{revtex4} 
\usepackage{amsmath,amssymb,amsfonts,bm}
\usepackage{graphicx,epstopdf} 
\usepackage{color}

\renewcommand{\d}{{\rm d}}
\newcommand{\iChEM}{{\it i}{\rm ChEM}}
\newcommand{\cf}{cf.\ }
\newcommand{\Sec}[1]{Sec.\,\ref{#1}}

\newcommand{\B}{\mbox{\tiny B}}
\newcommand{\tS}{\mbox{\tiny S}}
\newcommand{\T}{\mbox{\tiny T}}
\newcommand{\SB}{\mbox{\tiny SB}}
\newcommand{\BS}{\mbox{\tiny BS}}
\newcommand{\BB}{\mbox{\tiny BB}}

\newcommand{\dg}{\dagger}

\newcommand{\w}{\omega}
\newcommand{\wti}{\widetilde}
\newcommand{\la}{\langle}
\newcommand{\ra}{\rangle}
\newcommand{\La}{\big\la}
\newcommand{\Ra}{\big\ra}

\newcommand{\nl}{\nonumber \\}

\newcommand{\be}{\begin{equation}}
\newcommand{\ee}{\end{equation}}
\newcommand{\bea}{\begin{eqnarray}}
\newcommand{\eea}{\end{eqnarray}}
\newcommand{\bsube}{\begin{subequations}}
\newcommand{\esube}{\end{subequations}}
\newcommand{\Eq}[1]{Eq.\,(\ref{#1})}
\newcommand{\Eqs}[1]{Eqs.\,(\ref{#1})}
\newcommand{\Fig}[1]{Fig.\,\ref{#1}}

\begin{document}

\title{Theoretical formulations on thermodynamics of quantum impurity systems}

\author{Hong Gong} \thanks{Authors of equal contributions}
\author{Yao Wang} \thanks{Authors of equal contributions}
\author{Hou-Dao Zhang}
\author{Rui-Xue Xu}
\author{Xiao Zheng}
\author{YiJing Yan} \email{yanyj@ustc.edu.cn}

\affiliation{Hefei National
 Laboratory for Physical Sciences at the Microscale
 and Synergetic Innovation Center of Quantum Information and Quantum Physics
 and Collaborative Innovation Center of Chemistry for Energy Materials (\iChEM)
 and Department of Chemical Physics,
 University of Science and Technology of China,
 Hefei, Anhui 230026, China}

\date{August 19, 2020; Submit to PRB}

\begin{abstract}

 In this work, we put forward the theoretical foundation
toward thermodynamics of quantum impurity systems
measurable in experiments.
The theoretical developments involve
the identifications on two types of thermodynamic entanglement
free--energy spectral functions
for impurity systems that can be
either fermionic or bosonic or combined.
Consider further the thermodynamic limit in which
the hybrid environments satisfy the Gaussian--Wick's theorem.
We then relate the thermodynamic spectral functions
to the local quantum impurity systems spectral densities
that are often experimentally measurable.
Another type of inputs is the bare--bath
coupling spectral densities,
which could be accurately determined with
various methods.
Similar relation is also established
for the nonentanglement component that
exists only in anharmonic bosonic impurity systems.
For illustration, we consider the simplest
noninteracting systems, with focus on
the strikingly different characteristics
between the bosonic and fermionic scenarios.

\end{abstract}


\maketitle

\section{Introduction}
\label{thintro}


 Quantum impurity systems such as quantum dots
and nanostructured materials offer diversified functionalities,
where the strong correlations,
quantum entanglement, coherence and decoherence often play crucial roles.
The properties such as electronic and heat conductivity
of nano-materials
can be enhanced significantly compared to their bulk counterparts. \cite{Yon11131,Pek15118,Mil16011002}
These unique properties can be exploited
to design highly efficient molecular junctions and quantum devices. \cite{Gev967681,Hor132059,Skr144185,Fre17012146,Mer177678}
All these frontier developments need
to be guided by basic thermodynamic principles
in the quantum regime.\cite{Bin18}
The ever increasing capability in the exquisite
manipulations and detections
leads to quantum impurity systems
also ideal test beds for quantum physics.
However, can thermodynamics
be experimentally measurable,
particularly for quantum impurity systems?
This is an open question to be addressed.

 In this paper, we will exploit some basic relations
toward the above quest.
As Einstein remarked, ``\emph{Thermodynamics
is the only physical theory which I am convinced
will never be overthrown, within the framework of
applicability of its basic concepts}''.\cite{Ein79}
On this basis, we elucidate a set of universal relations
between mesoscopic quantum mechanics
and macroscopic thermodynamics.

 Physically, one can visualize a quantum impurity system
as a thermodynamic mixture, such as the widely used Anderson impurity model.
The total system--and--bath composite
Hamiltonian assumes the form of
$H_{\T} = H_{\tS}+ h_{\B}+ H_{\SB}$.
The last term denotes the  hybridization
between the local mesoscopic system and
a nonlocal macroscopic bath environment.
In the thermodynamics nomenclature, such a total composite
mixture at a given temperature $T$ constitutes
a closed system. It is in thermal contact with surrounding
heat reservoir to maintain the constant temperature scenario.

 We will see in \Sec{thsec2} there is a difference
between the fermionic and bosonic hybridization scenarios.
The former has only the entanglement component,
but the latter involves also the nonentanglement contribution.
Both these two thermodynamic components could be
experimentally measured.

 In \Sec{thsec3}, we present a unified theory
and relate the entanglement thermodynamics
to two types of spectral functions.
One is the entanglement free--energy spectral density,
with odd parity in frequency.
Another is the entanglement thermodynamic spectrum,
with even parity in frequency.
Interestingly, these two spectral functions possess
opposite parity, but equal--area in the half--side frequency region
of $\w\in[0,\infty)$.
While both are about equally accessible in experiments,
we would suggest the thermodynamic spectrum
be the choice.
 In Appendix, we analyze the universal
high--temperature thermodynamic behaviors.
We show the dramatic difference between
the fermionic and bosonic hybridization
scenarios.


 Consider further the theoretical formulations
with Gaussian bath environments,
where the Gaussian--Wick's theorem
is applicable.\cite{Kle09,Wei12,Yan05187}
This coupling bath model is rather commonly adopted
in various theories in quantum mechanics of open systems,
such as the path--integral influential functional 
formalism.\cite{Fey63118,Wei08195316,Muh08176403}
Its time--derivative equivalence,
the hierarchical equations of motion (HEOM) formalism,
either bosonic \cite{Tan89101,Tan906676,Xu05041103,Xu07031107}
or fermionic,\cite{Jin08234703}
is now a well--established method.\cite{Tan06082001,Tan20020901,Din11164107,%
Din12224103,Li12266403,Zhe121129,Zhe13086601,Ye16608}
The dissipaton equation of motion theory
is also developed.\cite{Yan14054105,Yan16110306,Zha18780,Wan20041102}
This is a statistical quasi--particle extension of the HEOM,
covering further the hybrid bath dynamics.
\cite{Yan14054105,Yan16110306,Zha18780,Wan20041102,Jin15234108,Jin20235144}

 We will show that the system--and--bath entanglement theory
with Gaussian environments\cite{Du20034102}
is intimately related to
the aforementioned entanglement free--energy spectral functions.
We will further extend this theory
to its treatment on the nonentanglement
thermodynamic component
that is generally nonzero for
bosonic quantum impurity systems.
We present the formulations with fermionic
and bosonic Gaussian environments
in \Sec{thsec4} and \Sec{thsec5}, respectively.
We conclude that the thermodynamic
hybridizing free--energy, either fermionic or bosonic,
can be completely determined with
the \emph{local impurity system} properties
and the nonlocal bath hybridization functions.
These two types of properties of quantum impurity systems
are in principle both experimentally measurable.
In \Sec{thnum}, we illustrate the results
on noninteracting systems
and show the remarkably distinct bosonic versus
fermionic characteristics.
We summarize this paper with \Sec{thconc}.


\section{Entanglement versus nonentanglement thermodynamics} %
\label{thsec2}

\subsection{Thermodynamic integral formalism}
\label{thsec2A}

  We will focus on
the free--energy change before and after hybridization:
\be\label{Ahyb_def}
 A_{\rm hyb}(T)\equiv A(T)-A_0(T).
\ee
This corresponds to $Z_\text{hyb}=e^{-\beta A_{\rm hyb}}=Z_{\T}/Z_{0}$,
with $\beta=1/(k_BT)$, whereas
$Z_{\T}=e^{-\beta A}= {\rm Tr} e^{-\beta H_{\T}}$
and
$Z_{0}=e^{-\beta A_0}=Z^{\tS}_0Z^{\B}_0=({\rm tr}_{\tS} e^{-\beta H_{\tS}})
({\rm tr}_{\B} e^{-\beta h_{\B}})$.
One can evaluate $Z_\text{hyb}=e^{-\beta A_{\rm hyb}}$
directly via the imaginary--time approaches,
such as the path--integral formalism\cite{Fun18040602}
and its influential functional derivative
equivalence.\cite{Tan14044114,Tan15144110,Kat18579}

 Alternatively, according to the Second Law,
one can relate the isotherm free--energy change to the
\emph{reversible work} performed on total composite mixture.
This results in $A_{\rm hyb}(T)$ the thermodynamic integral
formalism, with the varying system--bath
coupling strength as the integration parameter.\cite{Kir35300,Shu71413,Zon08041103,Zon08041104}
 To proceed, we write the total composite Hamiltonian
in the hybridization parameter $\lambda$--augmented
form,
\be\label{HT_lambda}
 H_{\T}(\lambda) = H_{\tS}+ h_{\B}+\lambda H_{\SB}.
\ee
A reversible process is now mathematically
described  with the smooth varying the hybridization parameter
from $\lambda=0$ to $\lambda=1$.
Denote $\hat\rho^{\rm eq}_{\T}(T;\lambda)
\equiv e^{-\beta H_{\T}(\lambda)}/Z_{\T}(\lambda)$.
The \emph{differential reversible work}
performed in $[\lambda,\lambda+\d\lambda]$
is then
\be\label{delW_rev}
 \delta w_{\rm rev}(\lambda)
 = {\rm Tr}[H_{\SB}\hat\rho^{\rm eq}_{\T}(T;\lambda)]\d\lambda.
\ee
We obtain the thermodynamic integral expression,\cite{Kir35300,Shu71413,Zon08041103,Zon08041104}
\be\label{Ahyb}
  A_{\rm hyb}(T) = \int^{1}_0\! \delta w_{\rm rev}(\lambda)
 = \int^{1}_0\! \frac{\d\lambda}{\lambda} \la H_{\SB}\ra_{\lambda},
\ee
with
\be\label{HSB_lambda}
 \la H_{\SB}\ra_{\lambda} \equiv
 {\rm Tr}[(\lambda H_{\SB})\hat\rho^{\rm eq}_{\T}(T;\lambda)].
\ee
This is just the $\lambda$--augmented
equivalence to the original $\la H_{\SB}\ra$ where $\lambda =1$.
Therefore, all methods on $\la H_{\SB}\ra$
would be readily applicable for thermodynamics.
These include the quantum Monte Carlo approach,%
\cite{Hir862521, Gul11349}
density matrix renormalization group,\cite{Whi922863,Vid03147902}
numerical renormalization group,\cite{Wil75773,Bul08395}
Green's function technique\cite{Kad62}
and multi-configuration time-dependent Hartree
method.\cite{Mey9073,Wan031289}
It is also noticed that the above formalism,
\Eq{Ahyb} with \Eq{HSB_lambda}, can be readily generalized
to transient thermodynamics problems.\cite{Gon20JCP}

\subsection{Entangled and nonentangled contributions}
\label{thsec2B}

In general the system--bath coupling
$H_{\SB}$ assumes a multiple--modes decomposition form,
with each mode being a product
of a system operator and a bath operator.
The hybridization pair of operators
can be either bosonic or fermionic.
The resultant $\la H_{\SB}\ra$ differs
in these two scenarios, as detailed below.

\subsubsection{Bosonic hybridization case}
\label{thsec2A1}

 The generic form of bosonic hybridization reads
\be\label{HSB_bose}
  H_{\SB} = \sum_{u} \hat Q_u\hat F_u,
\ \ \text{with} \ \
  [\hat Q_u,\hat F_v]=0.
\ee
Here, $\{\hat Q_u\}$ and $\{\hat F_u\}$
are Hermitian operators in the local impurity system
and the nonlocal bath subspaces, respectively.
Let $\delta\hat O \equiv \hat O - \la \hat O \ra$. We have
\be\label{HSB_uve_bose}
 \la H_{\SB}\ra = \sum_{u}\la\hat Q_u\ra\la\hat F_u\ra
  +\sum_{u} \la\delta\hat Q_u\delta\hat F_u\ra.
\ee
It involves both the uncorrelated and the nonlocally correlated sum terms.
Their $\lambda$--augmented counterparts
give rise the nonentanglement and entanglement free--energy
contributions to \Eq{Ahyb}, respectively.
The former reads
\be\label{Ahyb_nen}
 A_{\rm hyb}^{\rm nen}(T)
=\sum_u\int^{1}_0\frac{\d\lambda}{\lambda}
  \la \hat Q_u\ra_{\lambda}\la\hat F_u\ra_{\lambda},
\ee
where
\be\label{nen_Q_F_lambda}
\begin{split}
 \la\hat Q_u\ra_{\lambda}
&={\rm Tr}[\hat Q_u\hat\rho^{\rm eq}_{\T}(T;\lambda)],
\\
 \la\hat F_u\ra_{\lambda}
&={\rm Tr}[\lambda\hat F_u\hat\rho^{\rm eq}_{\T}(T;\lambda)].
\end{split}
\ee
Similarly, the entanglement free--energy contribution is
\be\label{Ahyb_en_bose}
 A^{\rm en}_{\rm hyb}\equiv A_{\rm hyb}-A_{\rm hyb}^{\rm nen}
=\sum_u\int^{1}_0\frac{\d\lambda}{\lambda}
  \la \delta\hat Q_u\delta\hat F_u\ra_{\lambda}.
\ee

\subsubsection{Fermionic hybridization case}
\label{thsec2A2}

 In contrast to \Eq{HSB_bose},
a fermionic hybridization usually reads
\be\label{HSB_fer}
  H_{\SB} = \sum_{u}
   \big(\hat a^{\dg}_{u}\hat F_{u} + \hat F^{\dg}_{u}\hat a_{u}\big),
\ \, \text{with}  \ \,
 \{\hat a^{\dg}_{u},\hat F_{v}\}=0.
\ee
Here, $\{\hat a_{u}\}$ and $\{\hat F_{u}\}$ are fermionic
operators in the local impurity system
and the nonlocal bath subspaces, respectively,
satisfying $\{\hat a^{\dg}_{u},\hat F_{v}\}=
\{\hat a_{u},\hat F_{v}\}=0$.
In practise, $\hat a_{u}$ ($\hat a^{\dg}_{u}$)
is the annihilation (creation) operator,
associated with the specified single--electron spin--orbital state
in the system subspace.
The nonlocal bath subspace operator $\hat F_{u}$
consists of a linear combination annihilation
operators in the bath subspace.
Apparently, $\la \hat a^{\dg}_{u}\ra
 = \la \hat F_{u} \ra = 0$, due to the underlying fermionic nature.
Therefore, the fermionic hybridization
is a pure entanglement event, with
\be\label{Ahyb_en_fer}
  A_{\rm hyb} = A^{\rm en}_{\rm hyb}
=\sum_{u}\int^{1}_0\frac{\d\lambda}{\lambda}
  \La \hat a^{\dg}_{u}\hat F_{u} + \hat F^{\dg}_{u}\hat a_{u}\Ra_{\lambda}.
\ee

 In the coming section, we will focus on
the entanglement thermodynamics
via \Eq{Ahyb_en_bose} and \Eq{Ahyb_en_fer}
for the bosonic and fermionic scenarios, respectively.
We will identity the spectral density
descriptions on entanglement thermodynamics
that is intimately related to the fluctuation--dissipation theorem
(FDT).\cite{Kle09,Wei12,Yan05187}
The nonentanglement $A^{\rm nen}_{\rm hyb}$, \Eq{Ahyb_nen},
which exists only for the bosonic case,
will be revisited in \Sec{thsec5}.

\section{Entanglement thermodynamic spectral functions}
\label{thsec3}

\subsection{Entanglement spectral density: Bosonic case}
\label{thsec3A}

 It is noticed that the entanglement thermodynamics
can be treated with the linear response theory,
without approximations.
In the following developments, we set the time variable
$t\geq 0$, unless specified further.
Consider the bosonic case, \Eq{HSB_bose}, where $[\hat Q_u, \hat F_u]=0$.
The relevant response function with
the \emph{system--and--bath symmetrization}
would be
\be\label{GSB_bose}
 \chi_{\SB}(t)=\frac{i}{2}\sum_u\!
  \La[\hat Q_u(t), \hat F_u(0)]+[\hat F_u(t), \hat Q_u(0)]\Ra.
\ee
This is a real and odd function. Define
\be\label{wti_chi_def}
 \wti\chi_{\SB}(\w) \equiv
\int^{\infty}_0\!\!\d t\, e^{i\w t}\chi_{\SB}(t)
\equiv \wti\chi^{(r)}_{\SB}(\w)+i\wti\chi^{(i)}_{\SB}(\w).
\ee
The real and imaginary parts,
$\wti\chi^{(r)}_{\SB}(\w)\equiv {\rm Re}\wti\chi_{\SB}(\w)$
and $\wti\chi^{(i)}_{\SB}(\w)\equiv {\rm Im}\wti\chi_{\SB}(\w)$,
satisfy $\wti\chi^{(r)}_{\SB}(-\w)=\wti\chi^{(r)}_{\SB}(\w)$ and
$\wti\chi^{(i)}_{\SB}(-\w)=-\wti\chi^{(i)}_{\SB}(\w)$,
respectively.
The related spectral density is given by
\be\label{calJ_bose}
 {\cal J}_{\SB}(\w)
\equiv \frac{1}{2i}\!\int^{\infty}_{-\infty}\!\!\d t\, e^{i\w t}
  \chi_{\SB}(t) ={\rm Im}\wti\chi_{\SB}(\w).
\ee

   Throughout this paper,
we set $\hbar=1$ for the unit of Planck constant.
Denote also $\hat O(t)\equiv e^{iH_{\T}t}
\hat O e^{-iH_{\T}t}$
and $\la(\,\cdot\,)\ra\equiv
{\rm Tr}_{\T}[(\,\cdot\,)e^{-\beta H_{\T}}]/Z_{\T}$,
with $H_{\T}\equiv H_{\T}(\lambda=1)$.
This defines $\chi_{\SB}(t)$ of \Eq{GSB_bose}.
Its $\lambda$--augmented counterpart, $\chi_{\SB}(t;\lambda)$,
is similar but with $H_{\T}(\lambda)$ of \Eq{HT_lambda}.
The resultant $\wti\chi_{\SB}(\w;\lambda)$
and ${\cal J}_{\SB}(\w;\lambda)$ are followed
as \Eqs{wti_chi_def} and (\ref{calJ_bose}).
The above convention follows that in
\Sec{thsec2} and is adopted throughout
this paper.

 To evaluate \Eq{Ahyb_en_bose},
we exploit the following identity that arises from the bosonic FDT,
\be\label{FDT_bose}
 \la H_{\SB}\ra^{\text{en}}_{\lambda} =
 \sum_u\La \delta\hat Q_u\delta\hat F_u\Ra_{\lambda}
=  \frac{1}{\pi}\!\int_{-\infty}^{\infty}\!\! \d\w
   \frac{{\cal J}_{\SB}(\w;\lambda)}{1-e^{-\beta\w}}.
\ee
One can then recast \Eq{Ahyb_en_bose} as
\be\label{A_varphi_bose}
 A_{\rm hyb}^{\rm en}(T) = -\frac{1}{\pi}\!
  \int_{-\infty}^{\infty}\!\! \d\w
   \frac{\varphi(\w)}{1-e^{-\beta\w}},
\ee
with the entanglement free--energy spectral density,
\be\label{varphi_def}
 \varphi(\w) \equiv -{\rm Im}\!\int^{1}_0
  \frac{\d\lambda}{\lambda}\,\wti\chi_{\SB}(\w;\lambda)
 =-\varphi(-\w).
\ee
The inclusion of a negative sign
to each of \Eqs{A_varphi_bose} and (\ref{varphi_def})
is made with the spontaneity convention.
The last identity highlights that
the entanglement free--energy spectral density,
$\varphi(\w)$, is \emph{antisymmetric}.
This symmetry is rooted at the chosen
\emph{system--and--bath symmetrization}
response function, \Eq{GSB_bose}.
It turns out to be instrumental to
the entanglement thermodynamic spectrum,
as detained in \Sec{thsec3C}.

\subsection{Entanglement spectral density: Fermionic case}
\label{thsec3B}

 It is noticed that convention theories
of fermionic impurity systems
go with the Green's function formalism,
involving anticommutators between two non-Hermition operators.
In relation to the evaluation of
\be\label{HSB_ave_fer}
 \la H_{\SB}\ra
= \sum_{u}\la \hat a_{u}\hat F^{\dg}_{u}+\hat F_{u}\hat a^{\dg}_{u}\ra,
\ee
the relevant Green's function is
\be\label{GSB_fer}
 G_{\SB}(t) =\sum_{u}
    \La\{\hat a_{u}(t),\hat F^{\dg}_{u}(0)\}
      +\{\hat F_{u}(t),\hat a^{\dg}_{u}(0)\}\Ra.
\ee
It satisfies
\be\label{GSBt_sym}
  G^{\ast}_{\SB}(t)=G_{\SB}(-t)
\text{\ \ and \ \ } G_{\SB}(0)=0.
\ee
Define [\cf\Eq{wti_chi_def}]
\be\label{wtiGw_def}
 \wti  G_{\SB}(\w)\equiv \int^{\infty}_{0}\!\!\d t\,
   e^{i\w t} G_{\SB}(t).
\ee
The related spectral density is given by
\be\label{calJ_fer}
 {\cal J}_{\SB}(\w) = \frac{1}{2}\!\int^{\infty}_{-\infty}\!\!\d t\,
   e^{i\w t} G_{\SB}(t)= {\rm Re}\,\wti  G_{\SB}(\w),
\ee
satisfying
\be\label{int_calJ_fer}
  \int^{\infty}_{-\infty}\!\! \d\w\, {\cal J}_{\SB}(\w)
= \pi G_{\SB}(t=0) = 0.
\ee
We obtain
\be\label{FDT_fer0}
 \la H_{\SB}\ra =\frac{1}{\pi}\!\int^{\infty}_{-\infty}\!\!\d\w\,
  \frac{{\cal J}_{\SB}(\w)}{1+e^{\beta\w}}
 =\frac{1}{\pi}\!\int^{\infty}_{-\infty}\!\!\d\w\,
  \frac{{\cal J}^{\rm odd}_{\SB}(\w)}{1+e^{\beta\w}}.
\ee
The first identity arises from the fermionic FDT.
The last one is the integrated equality,
in which ${\cal J}_{\SB}(\w)$
can be replaced by its odd function component,
\be\label{calJ_odd}
 {\cal J}^{\rm odd}_{\SB}(\w)
 \equiv \frac{1}{2}[{\cal J}_{\SB}(\w)-{\cal J}_{\SB}(-\w)]
=-{\cal J}^{\rm odd}_{\SB}(-\w).
\ee
The observations are as follows.
Consider the symmetry property of Fermi function,
\be\label{fer_fun}
  \frac{1}{1+ e^{\beta\w}}
=\frac{1}{2} -\frac{\sinh(\beta\w/2)}{2\cosh(\beta\w/2)}.
\ee
The first term, the constant $(1/2)$, does not contribute
to \Eq{FDT_fer0}, due to \Eq{int_calJ_fer}.
The second term is an odd function,
resulting in the integrated equality,
the second identity of \Eq{FDT_fer0}.

 Now it is readily to obtain \Eq{Ahyb_en_fer}
the expression,
\be\label{A_varphi_fer}
 A_{\rm hyb}(T)=A_{\rm hyb}^{\rm en}(T)
= -\frac{1}{\pi}\!\int_{-\infty}^{\infty}\!\! \d\w\,
   \frac{\varphi(\w)}{1+e^{\beta\w}}.
\ee
The involved free--energy spectral density reads
\be\label{varphi_fer}
 \varphi(\w)
= -\frac{1}{2}{\rm Re}\!\int^{1}_0 \frac{\d\lambda}{\lambda}
  \left[\wti G_{\SB}(\w;\lambda)-\wti G_{\SB}(-\w;\lambda)\right],
\ee
with $\varphi(-\w)=-\varphi(\w)$, the same parity as \Eq{varphi_def}.

\subsection{Entanglement free--energy spectrum
  and the equal--area theorem}
\label{thsec3C}

 Following \Eq{wti_chi_def}  and \Eq{wtiGw_def}, we have
\begin{align}
 \wti\chi_{\SB}(z;\lambda)
&\equiv \int^{\infty}_0\!\!\d t\, e^{izt}\chi_{\SB}(t;\lambda),
\label{wtichiz}
\\
 \wti G_{\SB}(z;\lambda)
&\equiv \int^{\infty}_0\!\!\d t\, e^{izt}G_{\SB}(t;\lambda).
\label{wtiGz}
\end{align}
These are analytical functions of $z$ in the upper--half plane.
It is noticed that the fermionic $\varphi(\w)$,
\Eq{varphi_fer}, engages both $\wti G_{\SB}(z;\lambda)$
and $\wti G_{\SB}(-z;\lambda)$.
The latter is an analytical functions of $z$ in
the lower--half plane.
Apparently, the above specified nature
of analytical functions preserves in
their $\lambda$--integrals.
We can then perform the frequency integration
in both \Eq{A_varphi_bose} and \Eq{A_varphi_fer},
by using the Cauchy's contour integration technique.
The poles inside the individual half--plane contour
integration arise only from the Matsubara
frequencies.

 The Cauchy's contour integration evaluations
on \Eq{A_varphi_bose} and \Eq{A_varphi_fer}
result in the unified expression,
\be\label{Ahyb_en_unified}
 A^{\rm en}_{\rm hyb}(T)
 = -\frac{\delta^{\pm}}{\beta}\vartheta(0)
   \pm\frac{2}{\beta}
   \sum_{n=1}^{\infty}\vartheta(\varpi^{\pm}_n).
\ee
Here, $\delta^{+}=0$ and $\delta^{-}=1$
for the fermionic and bosonic cases, respectively.
The second term engages the Matsubara frequencies,
$\{\varpi^{\pm}_n=(2n-1+\delta^{\pm})\pi/\beta;\,n=1,\cdots\infty\}$.

 Two remarkable implications arises from \Eq{Ahyb_en_unified}.
Firstly, it defines the so--called entanglement free--energy spectrum,
$\vartheta(\varpi)$, as follows.
By comparing between the bosonic \Eq{Ahyb_en_unified}
and \Eq{A_varphi_bose} with \Eq{varphi_def},
we obtain
\be\label{vartheta_def_bose}
 \vartheta(\varpi\geq 0)= -\int^{1}_{0}\!
  \frac{\d\lambda}{\lambda}\wti\chi_{\SB}(i\varpi;\lambda)
  =\vartheta^{\ast}(\varpi).
\ee
Its fermionic counterpart can be identified
by comparing between the fermionic \Eq{Ahyb_en_unified} and
\Eq{A_varphi_fer} with \Eq{varphi_fer}.
It results in
\be\label{vartheta_def_fer}
 \vartheta(\varpi\geq 0)
 = {\rm Im}\!\int^{1}_{0}\!
  \frac{\d\lambda}{\lambda}\wti G_{\SB}(i\varpi;\lambda).
\ee
In fact, $\wti\chi_{\SB}(i\varpi;\lambda)$
and $\wti G_{\SB}(i\varpi;\lambda)$
via \Eqs{wtichiz} and (\ref{wtiGz})
are the Laplace transformations, with $s=\varpi$,
on $\chi_{\SB}(t;\lambda)$ and $G_{\SB}(t;\lambda)$,
respectively.
Moreover, $\chi_{\SB}(t)$, \Eq{wti_chi_def},
is real, and so is the resultant $\wti\chi_{\SB}(i\varpi;\lambda)$,
as highlighted in the last identity of \Eq{vartheta_def_bose}.
For its use in \Eq{Ahyb_en_unified},
the individual $\vartheta(\varpi)$ above
is needed only for $\varpi\geq 0$.
Mathematically, we would have
\be\label{vartheta_ext}
 \vartheta(\varpi<0)\equiv \vartheta(|\varpi|),
\ee
since the Matsubara poles in
the upper/lower--half plane, $z=\pm i\varpi_n$, are
symmetric.

 Another remarkable property
is the \emph{equal area relation}:
\be\label{equal_area}
 \int_{0}^{\infty}\!\!\d\w\,\vartheta(\w)
=\int_{0}^{\infty}\!\!\d\w\,\varphi(\w).
\ee
This arises from the formal consideration
on the zero--temperature limit
to \Eq{Ahyb_en_unified}, resulting in an integral,
with the measure of $\varpi^{\pm}_{n+1}-\varpi^{\pm}_n=2\pi/\beta$.
Comparing the resultant $A^{\rm en}_{\rm hyb}$
with that of \Eq{A_varphi_bose} or \Eq{A_varphi_fer}
leads to \Eq{equal_area}.
Mathematically, one can view the
above zero--temperature limit as a method of
$\beta\rightarrow\infty$.
It is concerned only with the $\beta$ variable
in the Fermi/Boson function $f^{\pm}_{\beta}(\w)$.
Remarkably, the equal--area relation (\ref{equal_area})
remains hold for general
$\vartheta(\w)$ and $\varphi(\w)$,
with temperature $T$ dependence via parameters.

 Note that in \Eq{Ahyb_en_unified}
the first term exists only for the bosonic case.
It results from $1/(\beta\w)$,
the high--temperature term in the Bose function,
evaluated by using\cite{Kle09,Wei12,Yan05187}
\be\label{wtichiSB_0}
 \wti\chi_{\SB}(0)=\wti\chi^{(r)}_{\SB}(0)
=\frac{1}{\pi}\!\int^{\infty}_{-\infty}\!\!\d\w
 \frac{\wti\chi^{(i)}_{\SB}(\w)}{\w}.
\ee
The two identities arise from $\wti\chi^{(i)}_{\SB}(\w=0)=0$
and the Kramers--Kronig relation, respectively.

 It is worth re-emphasizing that
\Eq{Ahyb_en_unified} engages no contribution
from the constant component of Bose/Fermi function.
This is exact when the underlying spectral density $\varphi(\w)$
is antisymmetric.
While this requirement holds naturally for the bosonic case,
\Eq{varphi_def}, the possibility of anti-symmetrization
for the fermionic case
has to be scrutinized and
implemented, as \Eqs{calJ_fer}--(\ref{varphi_fer}).
Having  the entanglement thermodynamic spectrum,
$\vartheta(\varpi)$,
been properly defined in \Eqs{vartheta_def_bose}--(\ref{vartheta_ext}),
the equal--area relation (\ref{equal_area})
does hold for both the bosonic and fermionic cases.

 In Appendix, we present in detail
the asymptotic analysis on the universal
high--temperature thermodynamic behaviors.
Again, \Eq{Ahyb_en_unified} serves the
convenient starting point for this analysis.
We show the dramatic differences between
the fermionic and bosonic hybridization
scenarios, particularly in terms
of the entropy changes.

\section{Fermionic entanglement theory with Gaussian environments}
\label{thsec4}

\subsection{Opening remarks}
\label{thsec4A}

  Consider hereafter the theoretical formulations
with Gaussian bath environments.
This is concerned with the standard coupling bath model,
commonly used in open quantum systems.
In this model, the bath $h_{\B}$ constitutes a collection
of infinite noninteracting particles, either bosonic
or fermionic, whereas the hybrid bath
modes $\{\hat F_u\}$ are linear.
The simplicity arises here due to the underlying Gaussian--Wick's
theorem.\cite{Kle09,Wei12,Yan05187}
The influence of a Gaussian bath on an arbitrary system
is completely dictated by the interacting
spectral densities that are bare--bath subspace properties.

 It is noticed that Gaussian environments
go with the system--and--bath entanglement theory.\cite{Du20034102}
This theory relates the entangled response
functions, such as $\chi_{\SB}(t)$ of \Eq{GSB_bose},
to the local system properties,
with any given bare--bath spectral densities.
The corresponding relations for
the entanglement free--energy spectral functions,
$\varphi(\w)$ and $\vartheta(\varpi)$ of \Sec{thsec3},
will then be readily obtained.
We defer the bosonic theory to \Sec{thsec5},
where the existed nonentanglement $A^{\rm nen}_{\rm hyb}$, \Eq{Ahyb_nen},
will also be treated.

 In this section, we present
a comprehensive account on
the system--bath entanglement theory with
fermionic Gaussian coupling environments.
The total composite Hamiltonian reads
\be\label{HT_fer}
 H_{\T}=H_{\tS}+h_{\B}+\sum_{u}\big(
    \hat a^{\dg}_{u}\hat F_{u} + \hat F^{\dg}_{u}\hat a_{u}\big),
\ee
with
\be\label{hB_F_fer}
 h_{\B} = \sum_{k}\epsilon_{k} \hat d^{\dg}_{k}\hat d_{k}
\ \ \text{and} \ \
 \hat F_{u}=\sum_{k} t^{\ast}_{uk}\hat d_{k}.
\ee
Here, $\hat d^{\dg}_{k}$ and $\hat a^{\dg}_{u}$
($\hat d_{k}$ and $\hat a_{u}$)
are the creation (annihilation) operators
for an electron in the specified bath state
$|k\ra$ of energy $\epsilon_{k}$ and system $|u\ra$, respectively.
The coupling parameter $t_{uk}$ describes
an electron transfer between $|u\ra$ and $|k\ra$
of a same spin.
The local impurity system ($H_{\tS}$) is arbitrary,
containing often open--shell electrons with
strong Coulomb interactions,
under the influence of a fermionic coupling Gaussian environment.

 As \Sec{thsec3B}, we adopt the Green's function convention
for the fermionic theory.
Note that in general
\be\label{Gt_sym}
\begin{split}
 G_{AB}(t)&\equiv \la \{\hat A(t), \hat B^{\dg}(0)\}\ra=G^{\ast}_{BA}(-t),
\\
 G^{\ast}_{AB}(t)&=\la \{\hat A^{\dg}(t), \hat B(0)\}\ra.
\end{split}
\ee
Adopt also the convolution notation,
\be\label{conv_notation}
 f_1(t)\otimes f_2(t) \equiv \int^{t}_{0}\!\d\tau\, f_1(t-\tau)f_2(\tau).
\ee
Let $\wti f(\w)$ be the frequency resolution of $f(t)$,
such as \Eqs{wti_chi_def} and (\ref{wtiGw_def}),
\be\label{wti_fw_def}
 \wti f(\w) \equiv \int^{\infty}_0\!\!\d t\,e^{i\w t} f(t).
\ee
We have $\wti f(\w) =\wti f_1(\w)\wti f_2(\w)$
if $f(t)=f_1(t)\otimes f_2(t)$.

 Equation (\ref{hB_F_fer}) constitutes a Gaussian environment, with the
interacting bath spectral densities,
\be\label{Juvs_w}
 J_{uv}(\w)
 =\pi \sum_{k} t^{\ast}_{uk}t_{vk}
   \delta(\w-\epsilon_{k}).
\ee
Note that $u$ and $v$ appearing in pair carry a same spin,
due to the aforementioned nature of transferring coupling parameter.
From \Eq{hB_F_fer}, we have
\be\label{hatFBt}
 \hat F^{\B}_{u}(t)\equiv e^{ih_{\B}t}
\hat F_{us}e^{-ih_{\B}t}
=\sum_{k} t^{\ast}_{uk}e^{-i\epsilon_{k}t}\hat d_{k}.
\ee
Together with $\{\hat d_{k},\hat d^{\dg}_{k'}\}=\delta_{kk'}$,
we obtain
\be\label{FBt_F0}
 \big\{\hat F^{\B}_{u}(t),\hat F^{\dag}_{v}\big\}
=\sum_{k}t^{\ast}_{uk}t_{vk}e^{-i\epsilon_{k}t}
=g_{uv}(t),
\ee
with $g_{uv}(t)$ being the interacting bath Green's function
that is formally defined as [\cf\Eq{Gt_sym}]
\be\label{gt_def}
 g_{uv}(t) \equiv \La\{\hat F^{\B}_{u}(t),
      \hat F^{\B\dag}_{v}(0)\}\Ra_{\B}
= g^{\ast}_{vu}(-t).
\ee
One can then recast \Eq{Juvs_w} as
\be\label{Jw_def}
 J_{uv}(\w)
=\frac{1}{2}\big[\wti g_{uv}(\w)+\wti g^{\ast}_{vu}(\w)\big].
\ee
Note that in \Eq{gt_def},
both $\hat F^{\B}_{u}(t)\equiv e^{ih_{\B}t}
\hat F_{us}e^{-ih_{\B}t}$ [\Eq{hatFBt}]
and
$\la(\,\cdot\,)\ra_{\B}\equiv {\rm tr}_{\B}[(\,\cdot\,)
e^{-\beta h_{\B}}]/Z^{\B}_0$,
are defined in the bare--bath subspace,
rather than the total composite space.
In other words,
$\hat F^{\B}_{u}(t)\neq \hat F_{u}(t)$, except for $t=0$,
and $\la(\,\cdot\,)\ra_{\B}\neq \la(\,\cdot\,)\ra$,
unless it is a c-number in study.

\subsection{System--bath entanglement theory}
\label{thsec4B}

 The system--bath entanglement theory
is an input--output type of formalism.
The inputs for the fermionic theory
below are ${\bm g}(t)\equiv \{g_{uv}(t)\}$,
\Eq{gt_def},
and the local impurity Green's functions,
\be\label{Gt_local}
   G^{\tS\tS}_{uv}(t)\equiv
\la\{\hat a_{u}(t),\hat a^{\dg}_{v}(0)\}\ra.
\ee
The outputs are the nonlocal Green's functions,
\be\label{Gt_BB}
 G^{\BB}_{uv}(t)\equiv
\la\{\hat F_{u}(t),\hat F^{\dg}_{v}(0)\}\ra.
\ee
and
\be\label{Gt_SB}
\begin{split}
 G^{\SB}_{uv}(t)&\equiv \la\{{\hat a}_{u}(t),  \hat F^{\dg}_{v}(0)\}\ra,
\\
 G^{\BS}_{uv}(t)&\equiv \la\{{\hat F}_{u}(t),  \hat a^{\dg}_{v}(0)\}\ra.
\end{split}
\ee
Note that \Eq{GSB_fer} can be recast in terms of
these two quantities; see \Eq{GSBt_tr}.

 The theoretical development starts with
the evaluation on ${\hat F}_{u}(t)
\equiv e^{iH_{\T}t}{\hat F}_{u}e^{-iH_{\T}t}$,
via the formal solution to
\be\label{Heis_hatF}
 \dot{\hat F}_{u}=i[H_{\T},\hat F_{u}]=i[H_{\tS}+h_{\B}+H_{\SB},\hat F_{u}].
\ee
First of all, from \Eqs{HT_fer} and (\ref{hB_F_fer}), we obtain
\be\label{Heis_hat_d}
 \dot{\hat d}_k=-i\epsilon_{k}\hat d_k-i\sum_{v}t_{vk}\hat a_{v}.
\ee
Its solution reads
\be\label{hat_d_t}
 \hat d_k(t) = e^{-i\epsilon_{k}t}\hat d_k(0)
     -i\!\int^{t}_0\!\d\tau\, t_{vk}e^{-i\epsilon_k(t-\tau)}\hat a_{v}(\tau).
\ee
By applying it for $\hat F_{u}$ in \Eq{hB_F_fer},
followed by using \Eqs{hatFBt} and (\ref{FBt_F0}),
we obtain [\cf\Eq{conv_notation}]
\be\label{Langevin_eq}
 {\hat F}_{u}(t) = {\hat F}^{\B}_{u}(t) - i\sum_{v}
   g_{uv}(t)\otimes\hat a_{v}(t).
\ee
It together with $\{{\hat F}^{\B}_{u}(t),\hat a^{\dg}_{v}\}
=0$ via \Eq{hatFBt} result in
\be\label{GBS_t_element}
 G^{\BS}_{uv}(t)
 =-i\sum_{v'} g_{uv'}(t)\otimes G^{\tS\tS}_{v'v}(t).
\ee
The symmetry relation, \Eq{Gt_sym},
leads to further
\be\label{GSB_t_element}
  G^{\SB}_{uv}(t)=-i\sum_{v'} G^{\tS\tS}_{uv'}(t)\otimes g_{v'v}(t).
\ee
These identify the two output quantities of \Eq{Gt_SB}.
Moreover, \Eq{Langevin_eq} together with \Eq{gt_def}
result in
\be\label{GBB_t_element}
 G^{\BB}_{uv}(t)=g_{uv}(t)-i\sum_{v'} g_{uv'}(t)\otimes G^{\SB}_{v'v}(t).
\ee
Applying further \Eq{GSB_t_element} completes
the output quantity in \Eq{Gt_BB},
with the input functions,
\Eqs{gt_def} and (\ref{Gt_local}).
 In the matrix form, the above results are\cite{Du20034102}
\be\label{GBB_t}
 {\bm G}^{\BB}(t)= {\bm g}(t)
   -{\bm g}(t)\otimes {\bm G}^{\tS\tS}(t)\otimes {\bm g}(t),
\ee
and
\be\label{GSB_t}
\begin{split}
 {\bm G}^{\SB}(t)&=-i{\bm G}^{\tS\tS}(t)\otimes {\bm g}(t),
\\
 {\bm G}^{\BS}(t)&=-i{\bm g}(t)\otimes {\bm G}^{\tS\tS}(t).
\end{split}
\ee
In terms of frequency resolutions, \Eq{wti_fw_def},
they are
\be\label{GBB_w}
 \wti{\bm G}^{\BB}(\w)= \wti{\bm g}(\w)
   -\wti{\bm g}(\w)\wti{\bm G}^{\tS\tS}(\w)\wti{\bm g}(\w).
\ee
and
\be\label{GSB_w}
\begin{split}
 \wti{\bm G}^{\SB}(\w)&=-i\wti{\bm G}^{\tS\tS}(\w)\wti{\bm g}(\w),
\\
 \wti{\bm G}^{\BS}(\w)&=-i\wti{\bm g}(\w)\wti{\bm G}^{\tS\tS}(\w).
\end{split}
\ee
These two matrixes are of equal trace, with
\be\label{GSBw_tr}
 {\rm tr}\,\wti{\bm G}^{\SB}(\w)
 ={\rm tr}\,\wti{\bm G}^{\BS}(\w)
 = -i\,{\rm tr}[\wti{\bm g}(\w)\wti{\bm G}^{\tS\tS}(\w)].
\ee

\subsection{Thermodynamic spectral functions
 with fermionic Gaussian environments}
\label{thsec4C}

 It is noticed that, by using \Eq{Gt_SB}, we can recast
\Eq{GSB_fer} as
$G_{\SB}(t)={\rm tr}\,{\bm G}^{\SB}(t)+{\rm tr}\,{\bm G}^{\BS}(t)$.
Together with \Eq{GSBw_tr}, we obtain
\be\label{GSBt_tr}
 \wti G_{\SB}(\w) =-2i\,{\rm tr}[\wti{\bm g}(\w)\wti{\bm G}^{\tS\tS}(\w)].
\ee
This is the basis for revisiting various
entanglement thermodynamic spectral functions,
defined in \Sec{thsec3},
with the fermionic Gaussian  environments.

 Let us start with the form of $\wti G_{\SB}(\w;\lambda)$  via \Eq{GSBt_tr}.
As inferred from \Eq{gt_def},
$\wti{\bm g}(\w;\lambda)=\lambda^2 \wti{\bm g}(\w)$.
Therefore,
\be\label{wtiGSB_lambda_fer}
 \wti G_{\SB}(\w;\lambda)
=-2i\lambda^2\,{\rm tr}[\wti{\bm g}(\w)\wti{\bm G}^{\tS\tS}(\w;\lambda)].
\ee
The free--energy spectral functions,
\Eqs{varphi_fer} and (\ref{vartheta_def_fer}),
are then ($\varpi\geq 0$)
\begin{align}\label{varphi_fer_GE}
 \varphi(\w)
&=-\frac{1}{2}{\rm Im}\! \int^{1}_0\!\d\lambda^2
  \big[X(\w;\lambda)-X(-\w;\lambda)\big],
\\ \label{vartheta_fer_GE}
 \vartheta(\varpi)
&= -{\rm Re}\!\int^{1}_0\!\d\lambda^2\,
  {\rm tr}\big[\wti{\bm g}(i\varpi)\wti{\bm G}^{\tS\tS}(i\varpi;\lambda)],
\end{align}
where
\be\label{X_fer_GE}
 X(\w;\lambda)
\equiv {\rm tr}\big[\wti{\bm g}(\w)\wti{\bm G}^{\tS\tS}(\w;\lambda)].
\ee
Note that $\wti{\bm G}^{\tS\tS}(\w;\lambda)$ is
an even function of $\lambda$.

Let us repeat the two equivalent free--energy expressions,
\Eq{A_varphi_fer} and \Eq{Ahyb_en_unified}
for the fermionic case, as follows.
\be\label{A_final_fer}
 A_{\rm hyb}(T) =-\frac{1}{\pi}\!
  \int_{-\infty}^{\infty}\!\! \d\w
   \frac{\varphi(\w)}{1+e^{\beta\w}}
 =\frac{2}{\beta} \sum_{n=1}^{\infty}\vartheta(\varpi_n),
\ee
with $\{\varpi_n=(2n-1)\pi/\beta\}$,
the fermionic Matsubara frequencies.
Note that in the fermionic hybridization scenario,
$A_{\rm hyb}(T)=A^{\rm en}_{\rm hyb}(T)$
[\cf\Eq{Ahyb_en_fer}].
Moreover, the spectral density
$\varphi(\w)$ and the corresponding
spectrum  $\vartheta(\w)$
are of equal area [\Eq{equal_area}]
within $\w\in [0,\infty)$.

 Remarkably, the above formalism implies
the thermodynamics of quantum impurity
systems be measurable.
First of all, it is exact with
the Gaussian environment ansatz that is well satisfied
in the thermodynamic limit.
For quantum impurity systems,
such as quantum dots, one could continuously
adjust $\lambda$ the system--bath coupling strength.%
\cite{Foo15103112,Vel15410,Gam11030502,Gu171,Sca193011,Kaf1752333}
One can also measure the impurity spectral densities,\cite{Dam03473,Kol05085456}
resulting in the local Green's function,
$\wti{\bm G}^{\tS\tS}(\w;\lambda)$.
Another ingredient $\wti{\bm g}(\w)$ in \Eq{wtiGSB_lambda_fer}
is dictated by the bare--bath hybridization, $\{J_{uv}(\w)\}$ of \Eq{Jw_def},
that could be determined with various accurate methods.
The above formalism would imply that the
thermodynamics of fermionic quantum impurity systems
be measurable in experiments.

\section{Bosonic entanglement theory with Gaussian environments}
\label{thsec5}

\subsection{Nonentanglement contribution}
\label{thsec5A}

  It is worth reminding that in the bosonic case,
the nonentanglement component,
$A^{\rm nen}_{\rm hyb}\equiv A_{\rm hyb}-A^{\rm en}_{\rm hyb}$,
\Eq{Ahyb_nen}, is nonzero  in general,
except for noninteracting systems (\cf\Sec{thnumA}).
The bosonic theory presented below, in parallel to  \Sec{thsec4},
will naturally treat not only the entanglement
component,\cite{Du20034102}
but also the nonentanglement part.
This is concerned with relating the mean values of hybrid bath operators, $\{\la\hat F_u\ra\}$, to those local system dissipative modes,
$\{\la\hat Q_u\ra\}$; see \Eq{FtoQ}.

 Let us start with the bosonic counterpart to \Eq{FBt_F0},
\be\label{FBt_F0_bose}
  i[\hat F^{\B}_{u}(t),\hat F_v(0)]=
  i[\hat F^{\B}_{u}(t),\hat F^{\B}_v(0)]
\equiv  \phi_{uv}(t).
\ee
This commutator itself is a c-number and equals to the
bare--bath response function [\cf\Eq{gt_def}],
\be\label{phit_def}
 \phi_{uv}(t)
= i\La[\hat F^{\B}_{u}(t),\hat F^{\B}_{v}(0)]\Ra_{\B} .
\ee
As any response function between two Hermitian operators,
$\phi_{uv}(t)$ is real, satisfying $\phi_{vu}(-t)=-\phi_{uv}(t)$.
The bare--bath spectral density is given by [\cf\Eq{Jw_def}]
\be\label{Jw_def_bose}
 J_{uv}(\w)
= \frac{1}{2}\big[\wti\phi_{uv}(\w)-\wti\phi_{vu}(-\w)\big].
\ee
Denote for the use soon below
\be\label{eta_def}
 \eta_{uv} \equiv \int^{\infty}_0\!\!\d t\, \phi_{uv}(t)
=\wti\phi_{uv}(\w=0).
\ee

 The bosonic counterpart to \Eq{Langevin_eq} reads
\be\label{Langevin}
   \hat F_{u}(t)=\hat F^{\B}_{u}(t)
-\sum_{v}\phi_{uv}(t)\otimes\hat Q_v(t).
\ee
This immediately results in
\be\label{FtoQ}
 \la\hat F_{u}\ra = -\sum_{v}\eta_{uv} \la\hat Q_v\ra.
\ee
We can also obtain this interesting result
via the DEOM theory
that is exact with Gaussian environments.\cite{Yan14054105,Yan16110306}
The nonentanglement term in \Eq{HSB_uve_bose} becomes
\be\label{HSB_nent}
 \la H_{\SB}\ra^{\rm nen}
= \sum_{u}\la\hat Q_u\ra\la\hat F_u\ra
=-\sum_{uv}\eta_{uv}\la\hat Q_u\ra\la\hat Q_v\ra.
\ee
Note also that $\wti\phi_{uv}(\w;\lambda)=\lambda^2\wti\phi_{uv}(\w)$.
Consequently, \Eq{HSB_nent} leads to
the nonentanglement free--energy contribution,
\Eq{Ahyb_nen}, the final expression of
\be\label{Ahyb_nen_GE}
 A_{\rm hyb}^{\rm nen}(T)
=-\frac{1}{2}\sum_{uv} \eta_{uv}\!\int^{1}_0\!\d\lambda^2
  \la \hat Q_u\ra_{\lambda}\la\hat Q_v\ra_{\lambda}.
\ee
The individual $\la \hat Q_u\ra_{\lambda}$
is an even function of $\lambda$.

\subsection{Entanglement contribution}
\label{thsec5B}

 As specified earlier,  the entanglement thermodynamics
can be described in terms of thermodynamic
spectral functions, $\varphi(\w)$ [\Eq{varphi_def}]
and $\vartheta(\varpi)$ [\Eq{vartheta_def_bose}].
This description is rooted at
the system--and--bath symmetrized response function,
$\chi_{\SB}(t)$ of \Eq{GSB_bose}.
The inputs for its evaluation via the entanglement theory
are the bare--bath ${\bm\phi}(t)\equiv \{\phi_{uv}(t)\}$
and the local--system response functions [\cf\Eq{Gt_local}],
\be\label{chit_local}
   \chi^{\tS\tS}_{uv}(t)\equiv
i\La[\hat Q_{u}(t),\hat Q_{v}(0)]\Ra.
\ee
The outputs, especially those relevant to entanglement thermodynamics,
are the following two nonlocal response functions [\cf\Eq{Gt_SB}],
\be\label{chit_SB}
\begin{split}
 \chi^{\SB}_{uv}(t)&\equiv i\La[{\hat Q}_{u}(t),  \hat F_{v}(0)]\Ra,
\\
 \chi^{\BS}_{uv}(t)&\equiv i\La[{\hat F}_{u}(t),  \hat Q_{v}(0)]\Ra.
\end{split}
\ee
The bosonic system--bath entanglement theorem
reads\cite{Du20034102}
\be\label{chiSB_w}
\begin{split}
 \wti{\bm\chi}^{\SB}(\w)=-\wti{\bm\chi}^{\tS\tS}(\w)\wti{\bm\phi}(\w),
\\
 \wti{\bm\chi}^{\BS}(\w)=-\wti{\bm\phi}(\w)\wti{\bm\chi}^{\tS\tS}(\w).
\end{split}
\ee
These two matrixes are of equal trace, with
\be\label{chiSBw_tr}
 {\rm tr}\,\wti{\bm\chi}^{\SB}(\w)
 ={\rm tr}\,\wti{\bm\chi}^{\BS}(\w)
 = -{\rm tr}[\wti{\bm\phi}(\w)\wti{\bm\chi}^{\tS\tS}(\w)].
\ee
Apparently, \Eqs{chiSB_w} and (\ref{chiSBw_tr})
are the bosonic counterparts to \Eqs{GSB_w} and (\ref{GSBw_tr}),
respectively.

 Moreover, by using \Eq{chit_SB}, we
can recast \Eq{GSB_bose} as
$\chi_{\SB}(t)=\frac{1}{2}[{\rm tr}\,\wti{\bm\chi}^{\SB}(\w)
+{\rm tr}\,\wti{\bm\chi}^{\BS}(\w)]$,
resulting in
\be\label{wtichi_bose}
 \chi_{\SB}(t)=-{\rm tr}[\wti{\bm\phi}(\w)\wti{\bm\chi}^{\tS\tS}(\w)].
\ee
We obtain [\cf\Eq{wtiGSB_lambda_fer}]
\be\label{wtichi_lambda_bose}
 \wti\chi_{\SB}(\w;\lambda)
=-\lambda^2 {\rm tr}\big[\wti{\bm\phi}(\w)\wti{\bm\chi}^{\tS\tS}(\w;\lambda)].
\ee
The resultant \Eqs{varphi_def} and (\ref{vartheta_def_bose})
read ($\varpi \geq 0$)
\begin{align}\label{varphi_Xw_bose}
 \varphi(\w)
&= \frac{1}{2}\, {\rm Im}\!\int^{1}_0\!\d\lambda^2\,
  {\rm tr}\big[\wti{\bm\phi}(\w)\wti{\bm\chi}^{\tS\tS}(\w;\lambda)],
\\ \label{vartheta_Xw_bose}
 \vartheta(\varpi)
&= \frac{1}{2}\!\int^{1}_{0}\!\d\lambda^2\,
  {\rm tr}\big[\wti{\bm\phi}(i\varpi)\wti{\bm\chi}^{\tS\tS}(i\varpi;\lambda)].
\end{align}
Let us repeat \Eqs{A_varphi_bose} and (\ref{Ahyb_en_unified})
for the bosonic case below:
\begin{align}\label{bose_final}
 A_{\rm hyb}^{\rm en}(T)
&= -\frac{1}{\pi}\!
  \int_{-\infty}^{\infty}\!\! \d\w\,
   \frac{\varphi(\w)}{1-e^{-\beta\w}}
\nl&
 = -\frac{1}{\beta}\vartheta(0) -\frac{2}{\beta}
   \sum_{n=1}^{\infty}\vartheta(\varpi_n),
\end{align}
with $\{\varpi_n=2n\pi/\beta\}$ being the bosonic Matsubara frequencies.
See also \Eq{equal_area} for the equal area
of $\varphi(\w)$ and $\vartheta(\w)$, within $\w\in [0,\infty)$.

 Combining \Eq{Ahyb_nen_GE}, we obtain
the hybridizing free--energy,
$A_{\rm hyb}(T)=A_{\rm hyb}^{\rm nen}(T)+A_{\rm hyb}^{\rm en}(T)$,
in terms of the local properties,
$\la \hat Q_u\ra_{\lambda}$ and
$\wti\chi^{\tS\tS}_{uv}(\w;\lambda)$,
and the bare--bath $\wti\phi_{uv}(\w)$ or
$J_{uv}(\w)$ of \Eq{Jw_def_bose}.
The above formalism would imply that
the thermodynamics of bosonic quantum impurity
systems be also experimentally measurable.
Again, the key issues would be the tunability
with respect to the system--bath coupling strength.%
\cite{Foo15103112,Vel15410,Gam11030502,Gu171,Sca193011,Kaf1752333}

\section{Analytical results versus general remarks}
\label{thnum}

 In this section, we present
the concrete illustrations
with noninteracting systems.
However, we will also deduce some nontrivial
insides for thermodynamics of
arbitrary fermionic systems.
There are a number of striking different features
from their bosonic counterparts.
We thoroughly address those puzzles in study
with the underlying physical principles.

\subsection{Brownian oscillator systems}
\label{thnumA}

 The simplest noninteracting bosonic scenario is the
one--dimensional Brownian oscillator (BO) system.
This is concerned with a local harmonic oscillator
of frequency $\w_{\tS}$ and (dimensionless) coordinate $\hat q_{\tS}$,
embedded in a Gaussian environment.
The system--bath coupling is described with
$H_{\SB}=\hat q_{\tS}\hat F$; i.e., $\hat Q_{\tS}=\hat q_{\tS}$ here.
For the BO complex, $\la\hat q_{\tS}\ra=0$;
thus the nonentanglement $A^{\rm nen}_{\rm hyb}=0$
via \Eq{Ahyb_nen_GE}.

The BO system is analytically solvable.
The resultant local--system susceptibility function reads%
\cite{Kle09,Wei12,Yan05187} 
\be\label{chi_BO}
 \wti\chi_{\tS\tS}(\w)
=\frac{\w_{\tS}}{\w_{\tS}^2-\w^2-\w_{\tS}\wti\phi(\w)},
\ee
with $\wti\phi(\w)$ being the frequency resolution
on the interacting bath response function, $\phi(t)$ of \Eq{phit_def}.
Note that
\be\label{chiSB_BO}
  \wti\chi_{\SB}(\w)=-\wti\phi(\w)\wti\chi_{\tS\tS}(\w).
\ee
Moreover,
\be\label{BO_input}
 \wti\chi_{\SB}(\w;\lambda) = -\frac{\lambda^2\w_{\tS}\wti\phi(\w)}
  {\w_{\tS}^2-\w^2-\lambda^2\w_{\tS}\wti\phi(\w)}.
\ee
The resultant \Eqs{varphi_Xw_bose} and (\ref{vartheta_Xw_bose}), respectively,
are given by ($\varpi\geq 0$)
\begin{align}\label{varphi_BO}
 \varphi(\w)&=\frac{1}{2}\,{\rm Im}
 \big\{\!\ln [1+\phi(\w)\wti\chi_{\tS\tS}(\w)]\big\} ,
\\ \label{vartheta_BO}
 \vartheta(\varpi)
&=\frac{1}{2}\ln\big|1+\phi(i\varpi)\wti\chi_{\tS\tS}(i\varpi)\big|.
\end{align}
Interestingly, the free--energy spectral density,
$\varphi(\w)$, is just the half--\emph{phase} of
$1+\phi(i\varpi)\wti\chi_{\tS\tS}(i\varpi)=1-\wti\chi_{\SB}(\w)$,
whereas the Laplacian spectrum, $\vartheta(\varpi)$,
is the half--\emph{exponent}
of $|1-\wti\chi_{\SB}(i\varpi)|$.
It is worth noting that the above characteristics are
limited to noninteracting systems.
Equation (\ref{vartheta_BO}) reads explicitly
\be\label{vartheta_BO_final}
  \vartheta(\varpi\geq 0)
=\frac{1}{2}\ln\left|\frac{\w_{\tS}^2+\varpi^2}
  {\w_{\tS}^2+\varpi^2-\w_{\tS}\wti\phi(i\varpi)}\right|.
\ee
This is a continuous and even function; \cf\Eq{vartheta_ext}.

 On the other hand, $\varphi(\w)$ of \Eq{varphi_BO},
which is an odd function [$\varphi(-\w)=-\varphi(\w)$],
is related to the aforementioned phase property,
with the discontinuity (for $\w>0$):
\bsube\label{varphi_BO_final}
\begin{align}\label{varphi_BOa}
 &\varphi(\w\neq\w_{\tS})
=\dfrac{1}{2}\arg\left[
\frac{\w_{\tS}^2-\w^2}{\w_{\tS}^2-\w^2-\w_{\tS}\wti\phi(\w)}\right],
\\ \label{varphi_BOb}
 &\varphi(\w=\w_{\tS}-0^\text{\tiny +})
=-\dfrac{1}{2}\!\arg[\wti \phi(\w_{\tS})]+\dfrac{\pi}{2},
\\ \label{varphi_BOb}
 &\varphi(\w=\w_{\tS}+0^\text{\tiny +})
=-\dfrac{1}{2}\!\arg[\wti \phi(\w_{\tS})].
\end{align}
\esube
The discontinuity occurs at the BO frequency, $\w=\w_{\tS}$.
This is a feature of noninteracting systems.
The observed $\pi/2$ jump in $\varphi(\w)$,
at $\w=\w_{\tS}\pm 0^\text{\tiny +}$, arises from
the $\pi$--shift in the phase of
$[1+\wti\phi(\w)\chi_{\tS\tS}(\w)]$,
as implied in \Eq{varphi_BO}.

\subsection{Fermionic Brownian oscillator}
\label{thnumB}

 The simplest noninteracting fermionic case
is concerned with a spinless--dot
electronic system,
$\hat H_{\tS} = \epsilon_{\tS}\hat a^{\dag}\hat a$,
with a transfer coupling, $H_{\SB}=\hat a^{\dg}\hat F+\hat F^{\dg}\hat a$,
to a noninteracting electron reservoir environment.
Evaluate the Heisenberg equation of motion for the system,
resulting in $\dot{\hat a}(t)=-i\epsilon_{\tS}\hat a(t) - i\hat F(t)$.
By using \Eq{Langevin_eq}, we obtain
$\dot{\hat a}(t)=-i\epsilon_{\tS}{\hat a}(t)
 -g(t)\otimes\hat a(t)-i\hat F^{\B}(t)$,
and further
$\dot G_{\tS\tS}(t)=-i\epsilon_{\tS}G_{\tS\tS}(t)-g(t)\otimes G_{\tS\tS}(t)$.
Note that the initial value of $G_{\tS\tS}(t)=1$.
We obtain the well--known result of
\be\label{G_BO}
 \wti G_{\tS\tS}(\w)=\frac{i}{\w-\epsilon_{\tS}+i\wti g(\w)}.
\ee
Note that [\cf\Eq{GSBt_tr}]
\be\label{GSB_BO}
 \wti G_{\SB}(\w)=-2i\wti g(\w)\wti G_{\tS\tS}(\w).
\ee
The $\lambda$--augmented correspondence is then [\cf\Eq{BO_input}]
\be\label{Fer input}
 \wti G_{\SB}(\w;\lambda)
= \frac{2\lambda^2\wti g(\w)}{\w-\epsilon_{\tS}+i\lambda^2\wti g(\w)}.
\ee
Perform the thermodynamic integration and
obtain \Eqs{varphi_fer} and (\ref{vartheta_def_fer})
the expressions ($\varpi\geq 0$),
\begin{align}\label{spectral_fer1_BO}
 \varphi(\w) &= \frac{1}{2}\,{\rm Im}
  \bigg\{\!
   \ln\bigg[\frac{1-\wti g(\w)\wti G_{\tS\tS}(\w)}
           {1-\wti g(-\w)\wti G_{\tS\tS}(-\w)}
      \bigg]
  \bigg\},
\\ \label{spectral_fer3_BO}
 \vartheta(\varpi)
&= \ln\big|1-\wti g(i\varpi)\wti G_{\tS\tS}(i\varpi)\big|.
\end{align}
The last expression reads explicitly
\be\label{vartheta_fer_BO}
  \vartheta(\varpi\geq 0)
=\ln\bigg|\frac{i\varpi-\epsilon_{\tS}}
  {i\varpi-\epsilon_{\tS}+i\wti g(i\varpi)}\bigg|.
\ee
This is a continues and even function.

 In contrast, the phase, $\varphi(\w)$ of \Eq{spectral_fer1_BO},
is an odd function and discontinued at $\w=\pm\w_{\tS}$,
with $\w_{\tS}\equiv|\epsilon_{\tS}|$ and
the following explicit form (for $\w\geq 0$):
\bsube \label{varphi_fer_BO}
\begin{align}
 &\varphi(\w\neq\w_{\tS})
=\frac{1}{2}{\rm arg}\bigg[\frac{1-\wti g(\w)\wti G_{\tS\tS}(\w)}
{1-\wti g(-\w)\wti G_{\tS\tS}(-\w)}\bigg],
\\
&\varphi(\w=\w_{\tS}-0^\text{\tiny +})
=\frac{1}{2}\arg%
 \bigg[\frac{i\wti g(-\w_{\tS})}
   {2\w_{\tS}+i\wti g(\w_{\tS})}\bigg],
\\
&\varphi(\w=\w_{\tS}+0^\text{\tiny +})
=
\frac{1}{2}\arg\bigg[\frac{i\wti g(-\w_{\tS})}
   {2\w_{\tS}+i\wti g(\w_{\tS})}\bigg]-\frac{\pi}{2}.
\end{align}
\esube
Note that $\w_{\tS}\equiv |\epsilon_{\tS}|$.
Moreover, $\varphi(-\w)=-\varphi(\w)$,
the anti-symmetrization as implied in \Eq{varphi_fer},
whereas \Eq{varphi_fer_BO} describes only $\varphi(\w>0)$.
The discontinuity occurs at $\w_{\tS}\pm 0^\text{\tiny +}$,
at which $\varphi(\w)$ is subject to a $\pi/2$--phase jump.

\subsection{Numerical demonstrations and discussions}
\label{thnumC}

\begin{figure}
\includegraphics[width=0.38\textwidth]{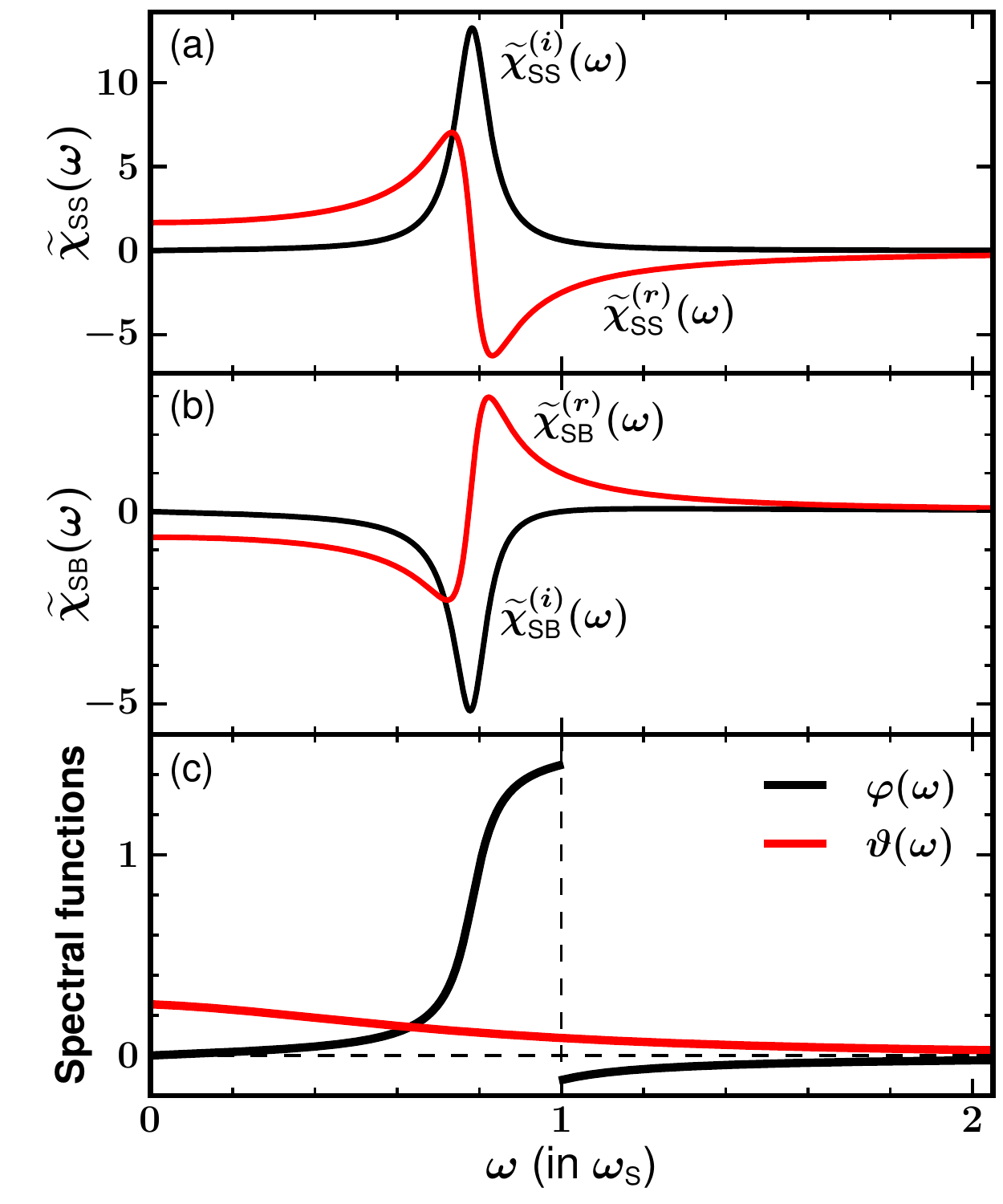}
\caption{Results on a bosonic BO
with a Drude environment.
(a) Local system $\wti\chi_{\tS\tS}(\w)$ [\Eq{chi_BO}],
 in unit of $\w^{-1}_{\tS}$, with the BO system frequency $\w_{\tS}$;
(b) Nonlocal $\wti\chi_{\tS\B}(\w)$ [\Eq{chiSB_BO}];
(c) Spectral functions, $\varphi(\w)$
[black; \Eq{varphi_BO_final}]
 and $\vartheta(\varpi=\w)$ [red; \Eq{vartheta_BO}].
See text for the environment parameters.
}
\label{fig1}
\end{figure}

\begin{figure}
\includegraphics[width=0.38\textwidth]{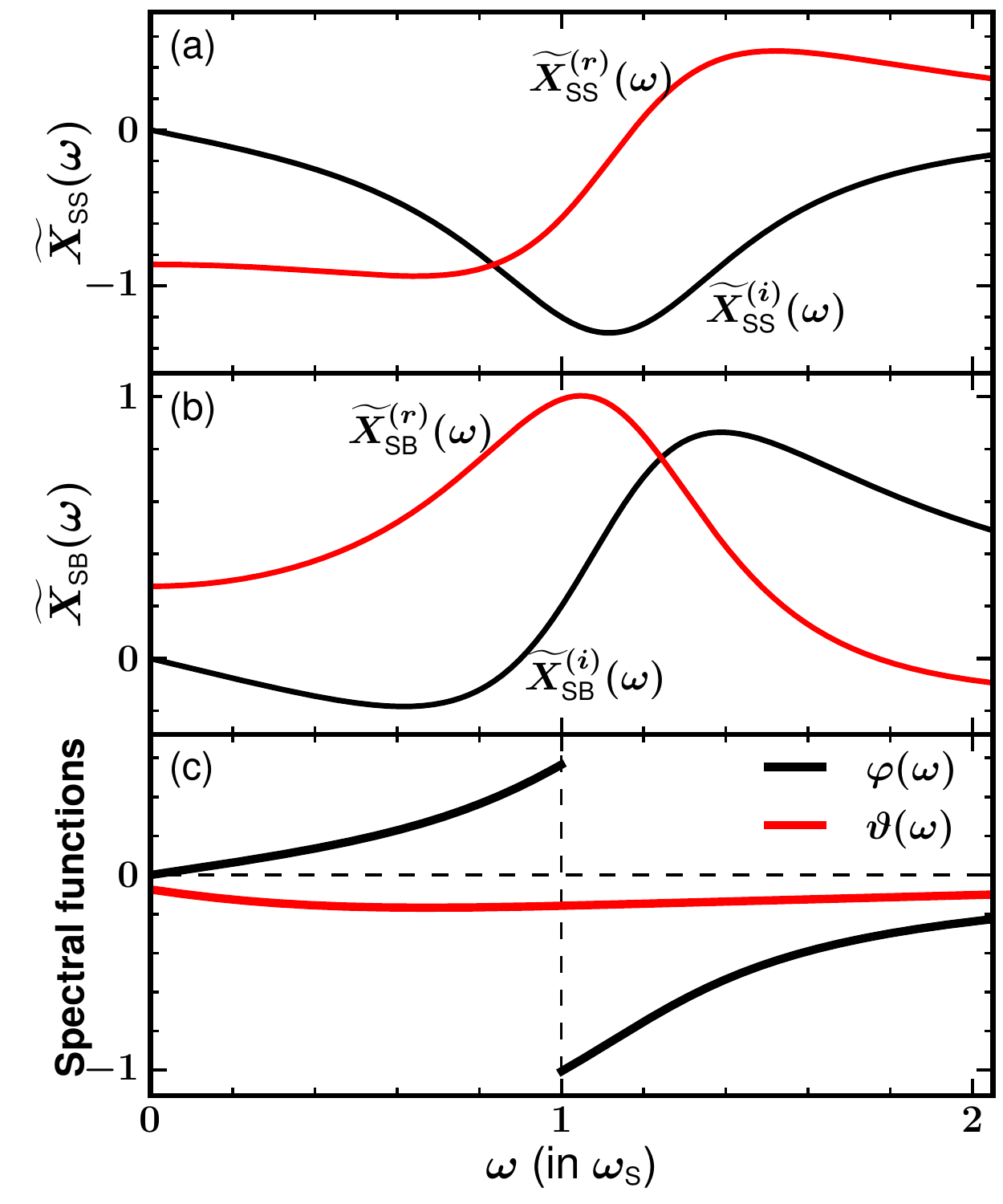}
\caption{The fermionic counterparts to those
in \Fig{fig1}.
(a) The resultant local $\wti X_{\tS\tS}(\w)$ [\cf\Eq{chiSSt_fer_def}],
in unit of $\w^{-1}_{\tS}$. The
system energy is negative, located at
$\epsilon_{\tS}=-\w_{\tS}$;
(b) Nonlocal $\wti X_{\tS\B}(\w)$ [\cf\Eq{chiSBt_fer_def}];
(c) Spectral functions,
 $\vartheta(\w)$ [black; \Eq{vartheta_fer_BO}]
 and $\varphi(\varpi=\w)$ [red; \Eq{varphi_fer_BO}].
}
\label{fig2}
\end{figure}

For their dictating
the thermodynamic spectral functions,
we would also like to show the response/Green's functions.
The bosonic case involves $\wti\chi_{\tS\tS}(\w)$ and $\wti\chi_{\SB}(\w)$,
whereas the fermionic case goes by $\wti G_{\tS\tS}(\w)$
and $\wti G_{\SB}(\w)$.
On the other hand, while $\chi_{\tS\tS}(t)$ and $\chi_{\SB}(t)$
are real, $G_{\tS\tS}(\w)$ and $G_{\SB}(t)$ are complex.
For the purpose of one-to-one comparison,
we set
\begin{align}\label{chiSSt_fer_def}
  X_{\tS\tS}(t) &= \frac{i}{2}\big[G_{\tS\tS}(t)-G_{\tS\tS}(-t)\big]
=-{\rm Im}G_{\tS\tS}(t).
\\ \label{chiSBt_fer_def}
  X_{\SB}(t) &= \frac{i}{2}\big[G_{\SB}(t)-G_{\SB}(-t)\big]
=-{\rm Im}G_{\SB}(t).
\end{align}
Consequently, the real parts of
$\wti\chi_{\tS\tS}(\w), \wti\chi_{\SB}(\w),
\wti X_{\tS\tS}(\w)$ and $\wti X_{\SB}(\w)$
are odd functions, whereas their imaginary parts
are even ones.
In fact, $\wti X^{(i)}_{\SB}(\w)={\cal J}^{\rm odd}_{\SB}(\w)$
of \Eq{calJ_odd} or the antisymmetrized
${\rm Im}\,X(\w;\lambda=1)$ of \Eq{X_fer_GE}.
Note also that the spectral functions,
$\varphi(\w)$ and $\vartheta(\w=\varpi)$,
are odd and even functions, respectively.

 Presented in \Fig{fig1} and \Fig{fig2} are
the calculated results on
the bosonic and fermionic cases, respectively.
Those even functions are in red
and the odd ones are in black.
Adopt for the demonstrations a Drude bath model,
\be\label{Drude_bath}
 \wti\phi(\w)=\frac{i\eta\gamma}{\w+i\gamma}=\wti g(\w),
\ee
with $\eta=0.4\w_{\tS}$  and $\gamma=4\w_{\tS}$,
for both the bosonic and fermionic BO systems.
The bosonic BO is of the frequency $\w_{\tS}$.
The fermionic BO is of the local on-site energy,
$\epsilon_{\tS}=-\w_{\tS}$, below the Fermi energy
of electronic bath reservoir.

\begin{figure}
\includegraphics[width=0.38\textwidth]{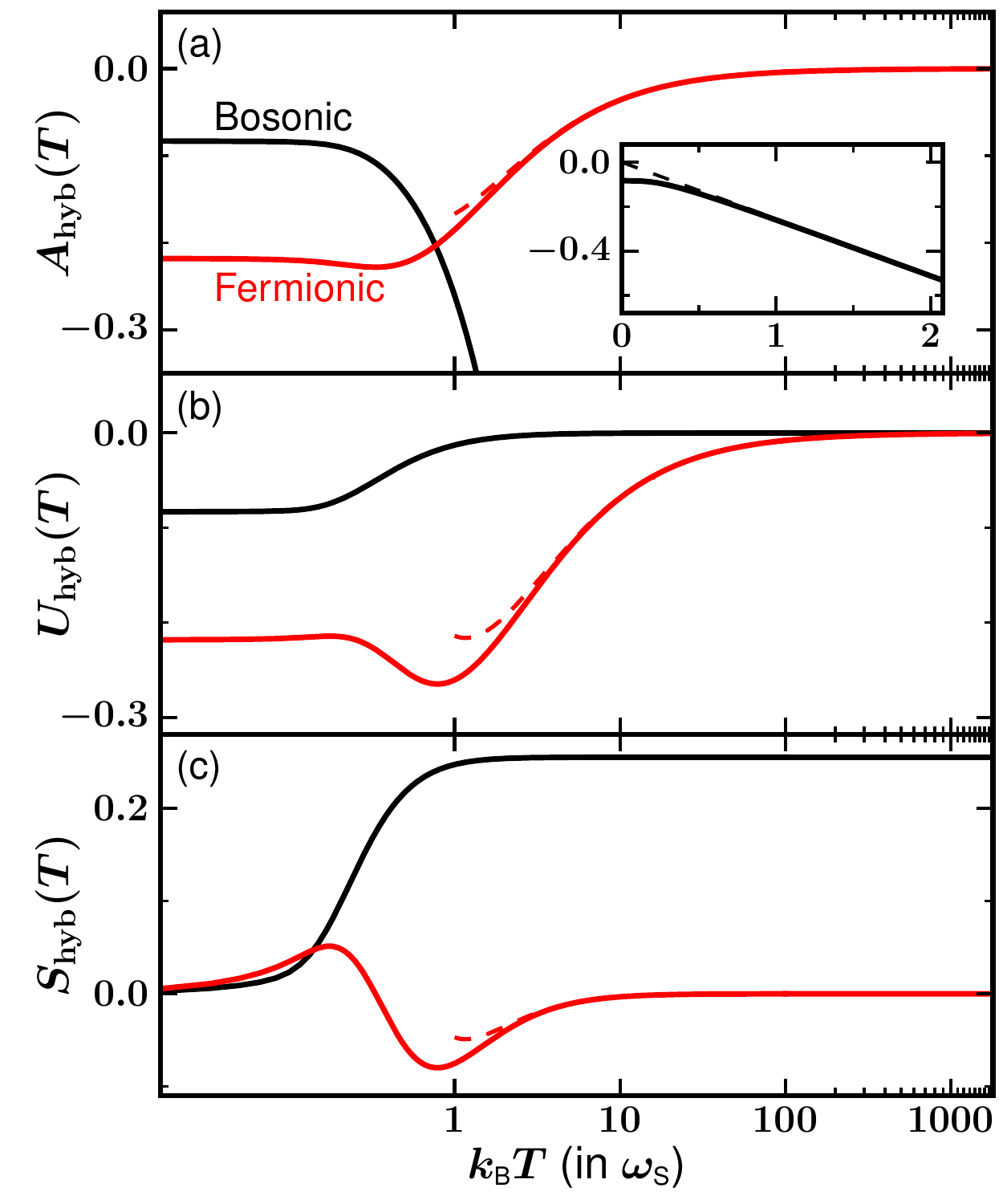}
\caption{Thermodynamic properties:
(a) Hybridization free--energy $A_{\rm hyb}(T)$, in unit of $\w_{\tS}$;
(b) Internal energy $U_{\rm hyb}(T)$, in unit of $\w_{\tS}$;
(c) Entropy $S_{\rm hyb}(T)$, in unit of $k_{B}$,
 for the bosonic (black) and fermionic (red) BO systems
 of \Fig{fig1} and \Fig{fig2}, respectively.
}
\label{fig3}
\end{figure}

 Reported in \Fig{fig3} are
 the hybridization (a) free--energy $A_{\rm hyb}(T)$,
(b) internal energy $U_{\rm hyb}(T)$, and (c)
entropy $S_{\rm hyb}(T)$,
for both the bosonic (black) and fermionic (red)
noninteracting systems.
The inset in \Fig{fig3}(a) depicts the linear--plot
on bosonic $A_{\rm hyb}(T)$. Included is also
its high--temperature
asymptotics, $-k_{B}T\vartheta(0)$ (dot);
see \Eq{bose_final} or \Eq{BO_highT_boseA}.
The observed $A_{\rm hyb}(T)<0$ indicates
the isotherm processes are spontaneous,
in both the bosonic and fermionic cases.

  We have numerically confirmed
the identities in individual
\Eq{A_final_fer} and \Eq{bose_final},
for both the bosonic and fermionic cases,
and also the equal--area relation (\ref{equal_area}).
The negative amplitude of the area
amounts to the value of $A_{\rm hyb}(T=0)$.
The resultant $S_{\rm hyb}(T=0)=0$
would also be general, for its agreeing with the Third Law.


  Interestingly, the BO systems,
bosonic versus fermionic,
show remarkably distinct behaviors in their
thermodynamic functions, especially
the hybridization entropy, $S_{\rm hyb}(T)$.
These will be elaborated together with
their high--temperature thermodynamic
characteristics, as follows.

The bosonic BO hybridization is
both energetically and  entropically favored,
with $U_{\rm hyb}(T)<0$ and
$S_{\rm hyb}(T)>0$, at all temperatures.
Moreover, in the high--temperature regime,
we have [\cf\Eq{highT_bose}]
\bsube\label{BO_highT_bose}
\begin{align} \label{BO_highT_boseA}
 \lim_{T\rightarrow\infty}\frac{A_{\rm hyb}(T)}{k_BT} &= -\vartheta(0),
\\ \label{BO_highT_boseS}
  S_{\rm hyb}(T\rightarrow\infty) &=k_{\B}\vartheta(0),
\\ \label{BO_highT_boseU}
  U_{\rm hyb}(T\rightarrow\infty) &= 0.
\end{align}
\esube
This is an ideal hybridization scenario.
In other words, in the high--temperature
limit, the bosonic BO mixtures
are ideal solutions in the elementary
physical chemistry.
The observed $U_{\rm hyb}(T)<U_{\rm hyb}(T\rightarrow\infty)$
reflects the quantum effect.
It seems as if \Eq{BO_highT_bose} were specialized
for noninteracting bosonic complexes,
where $A^{\rm en}_{\rm hyb}=A_{\rm hyb}$
and $\partial\vartheta/\partial T=0$.
However, we would argue, to the end of this section,
that \Eq{BO_highT_bose} be universal
for arbitrary bosonic systems in the limit
of $T\rightarrow \infty$.

It is worth noting that, for the bosonic noninteracting case,
$A_{\rm hyb}(T)$, $U_{\rm hyb}(T)$ and $S_{\rm hyb}(T)$
are all monotonic functions.
None of them shows the turnover behavior.
This would not be true for anharmonic systems,
such as the spin--boson complex, with these
three thermodynamic functions being evaluated
via a numerically accurate method.\cite{Gon20JCP}

 Turn to the fermionic case, the red--curves in \Fig{fig3}.
First of all, we have \Eq{fer_zeros} for the fermionic
case in general.
That is [\cf\Eq{BO_highT_bose}]
\be\label{BO_highT_fer}
 A_{\rm hyb}(\infty)=S_{\rm hyb}(\infty)=U_{\rm hyb}(\infty)=0.
\ee
These are universal fermionic relations
in the limit of $T\rightarrow \infty$.
We will discuss the physical picture
to the end of this section,
together with that of the bosonic \Eq{BO_highT_bose}.

Note that for noninteracting systems,
the thermodynamic spectrum is temperature--independent.
That is $\partial \vartheta(\varpi)/\partial\beta=0$.
The resultant \Eq{highT_PSD_app}, together with \Eq{highT_ferS},
read
\bsube\label{highT_PSD}
\begin{align}\label{AhighT_PSD}
 A_{\rm hyb}(T)
&\approx \frac{\kappa_a}{\xi_a} \varpi_a\vartheta(\varpi_a),
\\ \label{UhighT_PSD}
  U_{\rm hyb}(T)
&\approx
 -\frac{\kappa_a}{\xi_a}\varpi_a^2\vartheta'(\varpi_a),
\\ \label{ShighT_PSD}
 S_{\rm hyb}(T)
&\approx -\frac{\kappa_a}{\xi_a T}
  \big[\varpi_a\vartheta(\varpi_a)+\varpi_a^2\vartheta'(\varpi_a)\big].
\end{align}
\esube
Here $\kappa_a=3$, $\xi_a=\sqrt{12}$ and $\varpi_a=\xi_a/\beta$
[\Eq{kappa_PSD}], arising from the simplest Pad\'{e}
the [0/1]--approximant of Fermi function.\cite{Oza07035123,Hu10101106,Hu11244106}
Note that $\vartheta'(\varpi)\equiv \d\vartheta(\varpi)/\d\varpi$.
Included in \Fig{fig3} are also the high--temperature
approximants, \Eq{highT_PSD}, with the red--curves for $k_BT/\w_{\tS}>1$.
The accuracy is up to at least the order of ${\cal O}[(\beta\w_{\tS})^3]$,
as inferred from \Eq{PSD}.
Note that the local system energy is $\epsilon_{\tS}=-\w_{\tS}$,
below the Fermi energy of bath environment.
However, we can analytically prove that
the sign of $\epsilon_{\tS}$
does not affect $\vartheta(\varpi)$ in the present study.

 Strikingly, none of the fermionic $A_{\rm hyb}(T)$, $U_{\rm hyb}(T)$
and $S_{\rm hyb}(T)$ is monotonic.
In particular, the fermionic $S_{\rm hyb}(T)$ shows
a double--turnover characteristics.
The first one occurs in the entropically favored region,
with the maximum $S^{\rm max}_{\rm hyb}(T)>0$.
As temperature increases,
$S_{\rm hyb}(T)$ drops, getting into
the entropically unfavored region,
where the second turnover occurs,
with the minimum $S^{\rm min}_{\rm hyb}(T)<0$.
Afterward, it increases toward
$S_{\rm hyb}(T\rightarrow\infty)\rightarrow 0$.
It is noticed that the bosonic $A_{\rm hyb}(T)$, $U_{\rm hyb}(T)$
and $S_{\rm hyb}(T)$ consist only of turns,
occurring right in the temperature region
where seen the fermionic counterparts turnovers.
Therefore, we could attribute the observed
turnovers to the interplay between
system energy, thermal bath fluctuations
and the Pauli exclusion principle.

 To close this section, we would like to
address the physical picture behind
the bosonic \Eq{BO_highT_bose}
and fermionic \Eq{BO_highT_fer}.
Both comprise the universal relations
in the limit of $T\rightarrow\infty$.
In particular, the observed
$U_{\rm hyb}(T\rightarrow\infty) = 0$
in both bosonic and fermionic cases
agree perfectly with the
classical energy equipartition theorem.
The total number of degree of freedom is
invariant upon hybridization.
This observation explains also
the fermionic $S_{\rm hyb}(T\rightarrow\infty) = 0$.
The \emph{fermionic entropy equipartition theorem}
is the maximum qubit entropy
of $k_B\ln 2$ for each fermion.
This together with the aforementioned
energy equipartition theorem result
in further the fermionic $A_{\rm hyb}(T\rightarrow\infty) = 0$.
Interestingly, the fermionic entropy equipartition theorem
gives also rise to the observed
fermionic $S_{\rm hyb}(T) < 0$ in the high--temperature
regime, as it increases toward $S_{\rm hyb}(T\rightarrow\infty) = 0$.
In this regime, the Pauli exclusion results in
a lyophobic complex, with $U_{\rm hyb}(T) < 0$
and $S_{\rm hyb}(T) < 0$, prior to
the equipartition theorem takes the place.

\section{Concluding remarks}
\label{thconc}

  In summary, we have presented a comprehensive theory
of thermodynamics in the quantum regime.
Both the bosonic and fermionic hybridization scenarios
are considered.
We identify thermodynamic spectral functions,
together with the underlying relations (\Sec{thsec3} and Appendix).
By exploiting the system--bath entanglement theory,
we further relate the thermodynamic spectral functions
to experimental measurable quantities (\Sec{thsec4} and \Sec{thsec5}).

 It is noticed that there are two types of
thermodynamic spectral functions:
The free--energy spectral density, $\varphi(\w)$
[\Eq{varphi_def} or \Eq{varphi_fer}],
and the thermodynamic spectrum,
$\vartheta(\varpi)$
[\Eq{vartheta_def_bose} or \Eq{vartheta_def_fer}].
The former is defined in
the Fourier frequency domain with odd parity.
The latter is in
the Laplacian frequency domain with even parity.
Each of them completely characterizes the
entanglement thermodynamics properties.
Nevertheless, we would suggest the thermodynamic spectrum formalism
be the choice of convenience.

 We further show some remarkably different thermodynamic
characteristics between the bosonic and fermionic
noninteracting systems.
These provide the solid references for the future
studies on strongly correlated impurity complexes,
by using the general theories developed in this work.
 It is worth reemphasizing
the fact that both the bosonic \Eq{BO_highT_bose}
and the fermionic \Eq{BO_highT_fer}
are universal in the high--temperature limit.
We attribute these limiting results to
the equipartition theorem, as stipulated
to the end of \Sec{thnum}.

 It is noticed that the current
state--of--the--art devices available
for quantum simulation
include quantum dots, cold atoms/trapped ions,
superconducting circuits, etc.\cite{Foo15103112,Vel15410,Gam11030502,Gu171,Sca193011,Kaf1752333}
The established technologies on manipulating
such as the coupling conjunctions could be
exploited for the required thermodynamic
$\lambda$--integral here.
Therefore, the theoretical findings of this work
would constitute a crucial component for thermodynamics
in the quantum regime being measurable in experiments.

\begin{acknowledgments}
The support from
the Ministry of Science and Technology
(Nos.\ 2016YFA0400900, 2016YFA0200600 and 2017YFA0204904) 
the Natural Science Foundation of China
(Nos.\ 21633006, 21703225 and 21973086)
is gratefully acknowledged.
\end{acknowledgments}

\appendix*
\section{High--temperature regime:
Bosonic versus fermionic scenarios}
\label{thapp}

 This appendix presents the high--temperature
characteristics of entanglement
thermodynamic functions.
We will see there are dramatic differences
between the bosonic and fermionic hybridization cases.
Note that
$A^{\rm en}_{\rm hyb}(T)=U^{\rm en}_{\rm hyb}(T)-TS^{\rm en}_{\rm hyb}(T)$.
The hybridization entropy is
$S^{\rm en}_{\rm hyb}(T)=-\partial A^{\rm en}_{\rm hyb}(T)/\partial T$.

  Consider the bosonic case,
$A^{\rm en}_{\rm hyb}(T)$ of \Eq{Ahyb_en_unified}.
The first term there, $-k_{B}T\vartheta(0)$,
dominates the high--temperature properties.
We obtain
\bsube\label{highT_bose}
\begin{align}\label{highT_boseA}
   A^{\rm en}_{\rm hyb}(T)
&\stackrel{\text{high $T$}}{\longrightarrow}
  -k_BT\vartheta(0),
\\
  S^{\rm en}_{\rm hyb}(T)
&\stackrel{\text{high $T$}}{\longrightarrow}
  k_{\B}\vartheta(0)+k_{\B}T\frac{\partial \vartheta(0)}{\partial T},
\\
  U^{\rm en}_{\rm hyb}(T)
&\stackrel{\text{high $T$}}{\longrightarrow}
   k_{\B}T^2\frac{\partial\vartheta(0)}{\partial T}.
\end{align}
\esube
It is worth re-emphasizing that
$\vartheta(\varpi)$ depends in general
on temperature $T$. This dependence
is originated from the underlying response function,
$\chi_{\SB}(t)$ of \Eq{GSB_bose}.

 Turn to the fermionic case that does not have
the nonentanglement component; see \Eq{A_varphi_fer}.
We would have rather
\be\label{fer_zeros}
 0=A_{\rm hyb}(\infty)=S_{\rm hyb}(\infty)=U_{\rm hyb}(\infty).
\ee
These differ dramatically from the bosonic
counterparts in \Eq{highT_bose}.
The detailed derivations are as follows.

 To proceed, we consider \Eq{Ahyb_en_unified}
for the fermionic case, where
$\varpi_n=(2n-1)\pi/\beta$,
a suitable high--temperature approximant.
Let us start with
\be\label{A_fer_approximant}
 A_{\rm hyb}(T)=\frac{2}{\beta}
   \sum_{n=1}^{\infty}\vartheta(\varpi_n)
\approx \frac{\kappa}{\beta}\vartheta(\varpi_1),
\ee
This is the lowest Matsubara frequency based scheme,
with $\varpi_1=\pi/\beta$.
The resultant $\kappa = \pi^2/4$ will be identified
later, following the justifications,
\Eqs{kappa_def}--(\ref{MSD_highT_final}) and
comments there.
To the end of this appendix,
we will further propose
an optimal resum scheme;
see \Eqs{PSD}--(\ref{kappa_PSD}).

 Consider the temperature derivative on \Eq{A_fer_approximant},
which results in
\begin{align}\label{S_fer_approximant}
 S_{\rm hyb}(T)
&\approx
  -\kappa k_{B}\vartheta(\varpi_1)
-\frac{\kappa}{\beta}\pi k_{B}\vartheta'(\varpi_1)
\nl&\quad
 +\frac{\kappa}{T}\frac{\partial\vartheta(\varpi_1)}{\partial\beta}.
\end{align}
The last term arises from the intrinsic
temperature dependence of
$\vartheta(\varpi)$, which is originated from the underlying
Green's function, $G_{\SB}(t)$ of \Eq{GSB_fer}.
More precisely, $\partial\vartheta(\varpi)/\partial\beta\neq 0$,
whenever there is \emph{anharmonicity}.
Note also that
$\vartheta'(\varpi)\equiv\partial\vartheta/\partial\varpi$.
Equation (\ref{S_fer_approximant}) amounts to
\be\label{highT_ferS}
 S_{\rm hyb}(T)\approx
 -\kappa k_{B}\vartheta(\varpi_1)
 +U_{\rm hyb}(T)/T.
\ee
The first term is just $-A_{\rm hyb}(T)/T$,
with \Eq{A_fer_approximant}.

Let us express the hybridization free--energy
and internal energy in terms of (noting that $\varpi_1=\pi/\beta$)
\bsube\label{highT_fer}
\begin{align}\label{highT_ferA}
 A_{\rm hyb}(T)
&\approx \frac{\kappa}{\pi} \varpi_1\vartheta(\varpi_1),
\\ \label{highT_ferU}
  U_{\rm hyb}(T)
&\approx
 -\frac{\kappa}{\pi}\varpi_1^2\vartheta'(\varpi_1)
 +\kappa\frac{\partial\vartheta(\varpi_1)}{\partial\beta}.
\end{align}
\esube
The high--temperature limit
is then concerned with the three quantities,
$\varpi\vartheta(\varpi)$,
$\varpi^2\vartheta'(\varpi)$
and $\partial\vartheta(\varpi)/\partial\beta$,
in the $\varpi\rightarrow\infty$ regime.
The first two via \Eq{vartheta_def_fer}
are
\be\label{vartheta_def_app}
\begin{split}
 \varpi\vartheta(\varpi)
&= {\rm Im}\!\int^{1}_{0}\!
  \frac{\d\lambda}{\lambda} \varpi \wti G_{\SB}(i\varpi;\lambda),
\\
 \varpi^2\vartheta'(\varpi)
&= {\rm Im}\!\int^{1}_{0}\!
  \frac{\d\lambda}{\lambda} \varpi^2
   \frac{\partial}{\partial\varpi} \wti G_{\SB}(i\varpi;\lambda).
\end{split}
\ee
By using the asymptotics of
$\varpi e^{-\varpi t}\rightarrow 2\delta(t)$, we have
\[
 \varpi \wti G_{\SB}(i\varpi)
= \varpi\!\int^{\infty}_{0}\!\!\d t\, e^{-\varpi t} G_{\SB}(t)
\stackrel{\varpi\rightarrow\infty}{\longrightarrow}
 G_{\SB}(t=0).
\]
Moreover, by using
$\varpi^2 e^{-\varpi t}\rightarrow -2\dot\delta(t)$,
we have
\begin{align*}
 \varpi^2\frac{\d}{\d\varpi}\wti G_{\SB}(i\varpi)
&=-\!\int^{\infty}_0\!\!\d t\, (\varpi^2 e^{-\varpi t}) [tG_{\SB}(t)]
\nl&
 \stackrel{\varpi\rightarrow\infty}{\longrightarrow}
 2\!\int^{\infty}_0\!\!\d t\,\dot\delta(t)[tG_{\SB}(t)]
\nl&= -G_{\SB}(t=0).
\end{align*}
We can therefore write the limiting
values of \Eq{vartheta_def_app} as\
\be\label{ferHTfinal_1}
 \lim_{T\rightarrow\infty} \varpi\vartheta(\varpi)
=-\lim_{T\rightarrow\infty} \varpi^2\vartheta'(\varpi),
\ee
with
\be\label{ferHTfinal_2}
 \lim_{T\rightarrow\infty} \varpi\vartheta(\varpi)
={\rm Im}\!\int^{1}_{0}\!
  \frac{\d\lambda}{\lambda} G_{\SB}(t=0;\lambda)
=0.
\ee
The last identity follows \Eq{int_calJ_fer}.
 Moreover, for fermionic hybrid systems
in the high--temperature limit,
the intrinsic temperature--dependence of
$\vartheta(\varpi)$ via the Green's function
would be saturated. In other words,
\be\label{ferHTfinal_3}
 \lim_{\beta\rightarrow 0}
 \frac{\partial\vartheta(\varpi)}{\partial\beta} = 0.
\ee
 By applying \Eqs{ferHTfinal_1}--(\ref{ferHTfinal_3})
for the $T\rightarrow\infty$ limiting values
of \Eq{highT_fer}, we obtain immediately
all identities in \Eq{fer_zeros}.

 We are now in the position to elaborate the parameter, $\kappa=\pi^2/4$,
exploited in the second identity of \Eq{A_fer_approximant}.
Let us revisit this identity, with
the parameter $\kappa$ the formal expression,
\be\label{kappa_def}
 \kappa = \frac{2}{\vartheta(\varpi_1)}
  \sum_{n=1}^{\infty}\vartheta(\varpi_n).
\ee
The involving $\{\varpi_n\}$ are the
Matsubara frequencies, arising from
the Fermi function expansion,
\be\label{fermi_MSD}
 \frac{1}{1+e^{\beta\w}}
=\frac{1}{2}-2\sum_{n=1}^{\infty}
 \frac{\w/\beta}{\w^2+\varpi^2_{n}}.
\ee
Consider then the high--temperature approximation,
\be\label{MSD_highT_1}
 \frac{\w/\beta}{\w^2+\varpi^2_{n}}
\approx
 \frac{\beta\w}{(2n-1)^2\pi^2} + {\cal O}\big[(\beta\w)^3\big].
\ee
Therefore,
\be\label{MSD_highT_2}
\sum_{n=1}^{\infty} \frac{\w/\beta}{\w^2+\varpi^2_{n}}
\approx
 \frac{\beta\w}{\pi^2}\sum_{n=1}^{\infty} \frac{1}{(2n-1)^2}
=\frac{\beta\w}{\pi^2} \frac{\pi^2}{8}.
\ee
Together with $\frac{\w/\beta}{\w^2+\varpi^2_{1}}
\approx \frac{\beta\w}{\pi^2}$ via \Eq{MSD_highT_1},
we obtain
\be\label{MSD_highT_final}
 \sum_{n=1}^{\infty} \frac{\w/\beta}{\w^2+\varpi^2_{n}}
\approx \frac{\pi^2}{8}
 \frac{\w/\beta}{\w^2+\varpi^2_{1}}.
\ee
 The factor of $\pi^2/8$ represents
the ratio between the linear--order expansion
and the lowest Matsubara frequency expansion.
Remarkable, whenever the high--temperature asymptotic behaviors
are concerned with,
this ratio is  generic and
transferable to such as \Eq{kappa_def},
where $\kappa = 2\pi^2/8 = \pi^2/4$.
This is the value of $\kappa$ in
\Eqs{A_fer_approximant}--(\ref{highT_fer}).

 For a close comparison with an optimized
scheme [\cf\Eq{PSD}], we summarize the above
high--temperature approximant, in terms
of the Fermi function, \Eq{fermi_MSD}.
That is
\be\label{fermi_MSD_highT}
 \frac{1}{1+e^{\beta\w}}
\approx \frac{1}{2}-\frac{\pi^2}{4}
 \frac{\w/\beta}{\w^2+(\pi/\beta)^2}
 +{\cal O}\big[(\beta\w)^3\big].
\ee
The value of $\varpi_1=\pi/\beta$ is substituted explicitly.

 On the other hand, it is well--known
the best sum--over--poles expansion
for Bose or Fermi functions
is the Pad\'{e} spectrum decomposition scheme.\cite{Oza07035123,Hu10101106,Hu11244106}
For the high--temperature asymptotics,
it requires only the lowest--order Pad\'{e} $[0/1]$
approximant that reads
\be\label{PSD}
 \frac{1}{1+e^{\beta\w}}
\approx \frac{1}{2} -
 \frac{3\w/\beta}{\w^2+(\sqrt{12}/\beta)^2}
 + {\cal O}\big[(\beta\w)^5\big].
\ee
Its advantage over \Eq{fermi_MSD_highT} is clearly evident.
The involved single pole--related frequency
is no longer $\varpi_1=\pi/\beta$,
but rather $\varpi_a\equiv\sqrt{12}/\beta$.
The associated parameter is now $\kappa_{a} = 3$.
More important, \Eq{PSD} suggests \Eq{highT_fer}
be modified with
\bsube\label{highT_PSD_app}
\begin{align}\label{AhighT_PSD_app}
 A_{\rm hyb}(T)
&\approx \frac{\kappa_a}{\xi_a} \varpi_a\vartheta(\varpi_a),
\\ \label{UhighT_PSD_app}
  U_{\rm hyb}(T)
&\approx
 -\frac{\kappa_a}{\xi_a}\varpi_a^2\vartheta'(\varpi_a)
 +\kappa_a\frac{\partial\vartheta(\varpi_a)}{\partial\beta},
\end{align}
\esube
where
\be\label{kappa_PSD}
 \kappa_a=3, \ \ \xi_a= \sqrt{12} \ \
\text{and} \ \ \varpi_a = \xi_a/\beta.
\ee


\end{document}